%% file: main.tex
\documentclass[sn-mathphys,Numbered]{sn-jnl}

\usepackage{graphicx}%
\usepackage{multirow}%
\usepackage{amsmath,amssymb,amsfonts}%
\usepackage{amsthm}%
\usepackage{mathrsfs}%
\usepackage[title]{appendix}%
\usepackage{xcolor}%
\usepackage{textcomp}%
\usepackage{manyfoot}%
\usepackage{booktabs}%
\usepackage{algorithm}%
\usepackage{algorithmicx}%
\usepackage{algpseudocode}%
\usepackage{listings}%
\usepackage[utf8]{inputenc}
\usepackage{graphicx, nicefrac}
\usepackage{bm}
\usepackage{amsmath,MnSymbol,nicefrac}
\usepackage{lineno} 

\theoremstyle{thmstyleone}%
\theoremstyle{thmstyletwo}%

\theoremstyle{thmstylethree}%

\newcommand{\Cov}{\mathrm{Cov}}
\newcommand{\PCov}{\mathrm{PCov}}


\begin{document}

\title[Attosecond Delays in X-ray Molecular Ionization]{Attosecond Delays in X-ray Molecular Ionization}

\input{authors.tex}
\date{}

\abstract{

The photoelectric effect is not truly instantaneous, but exhibits attosecond delays that can reveal complex molecular dynamics.
Sub-femtosecond duration light pulses provide the requisite tools to resolve the dynamics of photoionization. 
Accordingly, the past decade has produced a large volume of work on photoionization delays following single photon absorption of an extreme ultraviolet~(XUV) photon. 
However, the measurement of time-resolved core-level photoionization remained out of reach.  
The required x-ray photon energies needed for core-level photoionization were not available with attosecond tabletop sources.  
We have now measured the x-ray photoemission delay of core-level electrons, and here report unexpectedly large delays, ranging up to 700 attoseconds in NO near the oxygen \textit{K}-shell threshold.
These measurements exploit attosecond soft x-ray pulses from a free-electron laser~(XFEL) to scan across the entire region near the \textit{K}-shell threshold. 
Furthermore, we find the delay spectrum is richly modulated, suggesting several contributions including transient trapping of the photoelectron due to shape resonances, collisions with the Auger-Meitner electron that is emitted in the rapid non-radiative relaxation of the molecule, and multi-electron scattering effects.
The results demonstrate how x-ray attosecond experiments, supported by comprehensive theoretical modelling, can unravel the complex correlated dynamics of core-level photoionization. 

} 

\maketitle

\paragraph{Introduction}
Understanding complex multi-electron interactions is a frontier scientific challenge since electron-electron interactions~(or correlations) play a fundamental role in determining the properties of matter.
The photoelectric effect~(or photoionization) in isolated atoms or molecules is inherently a multi-electron process.
Even if only one electron is removed from a molecule, the electronic wavefunction of the residual ion will rearrange during and after the interaction with the ionizing field. 
This effect is especially pronounced when the electron is removed from a core-level orbital of a molecular system. 

In this work, we use attosecond x-ray pulses to measure the temporal retardation of photoemission between electrons emitted from the oxygen and nitrogen $K$-shells of nitric oxide.
The photoelectric effect was long treated as an instantaneous process, because the extreme timescales associated with photoionization dynamics were inaccessible to experimental science.
However, the development of attosecond duration light pulses, has made it possible to achieve the requisite time resolution to resolve the dynamics of photoionization~\cite{schultze_delay_2010, nobelprizephys2023}. 
Time-domain measurements of photoemission have provided rich information on electron correlation effects in the system being ionized~\cite{ossiander_attosecond_2017,isinger_photoionization_2017,mansson_double_2014,biswas_probing_2020,haessler_phase-resolved_2009,gruson_attosecond_2016,kotur_spectral_2016,kamalov_electron_2020,cirelli_anisotropic_2018,peschel_attosecond_2022}, which is inaccessible to other electronic observables, such as electron binding energy, partial-photoionization cross-section, or photoelectron angular distribution.
The photoemission time delay can be related to the kinetic energy dispersion of the phase of the dipole matrix element for x-ray photoionization~\cite{pazourek_attosecond_2015}, which enables the complete measurement~(amplitude and phase) of a fundamental quantum phenomenon, the photoelectric effect.
Moreover, we find that the inclusion of electron correlation effects are critical for accurately modeling our experimental results.
In doing so, we demonstrate that $K$-shell photoemission delays offer a sensitive experimental probe of correlated electron motion in multi-electron systems.

Previous measurements of photoemission delays has employed methods such as RABBITT~\cite{veniard_phase_1996, paul_observation_2001,klunder_probing_2011} and laser streaking~\cite{kienberger_atomic_2004,schultze_delay_2010}. 
Initially used to study photoemission processes in atoms~\cite{schultze_delay_2010,klunder_probing_2011,dahlstrom_theory_2013}, these techniques have also been extended to molecular systems~\cite{biswas_probing_2020,haessler_phase-resolved_2009,kamalov_electron_2020,vos_orientation-dependent_2018,huppert_attosecond_2016,nandi_attosecond_2020,heck_attosecond_2021}.
These studies have lead to a deeper understanding of photoionization, particularly ionization in the vicinity of continuum structures~\cite{haessler_phase-resolved_2009,kamalov_electron_2020,vos_orientation-dependent_2018,huppert_attosecond_2016,nandi_attosecond_2020,ossiander_attosecond_2017,isinger_photoionization_2017,mansson_double_2014,biswas_probing_2020,gruson_attosecond_2016,kotur_spectral_2016,cirelli_anisotropic_2018,heck_attosecond_2021,peschel_attosecond_2022}.
To date, a significant limitation of these measurements has been the sparsity at which the electron kinetic-energy-dependent photoionization delay can be probed. 
In typical experiments only a few~(3-5) kinetic energy points can be collected, and this limits the ability to compare these results with theory. 
However with the advent of tunable attosecond-XFEL sources~\cite{duris_tunable_2020}, it is now possible to tune across a large range of electron kinetic energies.
To this end, we employ attosecond angular streaking~\cite{itatani_attosecond_2002,hartmann_attosecond_2018,eckle_attosecond_2008,duris_tunable_2020,li_attosecond_2022} using x-ray pulses from a free-electron laser to extend measurements of photoemission time-delays to core-level~($K$-shell) electrons.

\paragraph{Attosecond angular streaking}
In angular streaking an ionizing attosecond x-ray pulse is overlapped with a circularly-polarized, long wavelength~(infrared) laser field. 
The infrared~(IR) dressing field maps the temporal profile of the x-ray photoemission to the final momentum of the photoelectrons.
This mapping is similar to the principle of a time-resolving streak camera~\cite{tsuchiya_advances_1984} and takes place \textit{via} the so-called streaking interaction~\cite{bradley_direct_1971, kienberger_atomic_2004}. 
In a semi-classical approximation, the final momentum of an ionized electron measured at a detector is described (in atomic units) by
\begin{linenomath}\begin{equation}
    \vec{p}(t\rightarrow\infty) = \vec{p}_0 + e\vec{A}(t_0),
    \label{eqn:streak_mom}
\end{equation}\end{linenomath}
where $\vec{A}(t_0)=-\int_{-\infty}^{t_0}\vec{\mathcal{E}}_L(t^{\prime}) dt^{\prime}$ is the vector potential of the circularly polarized laser field,~$\mathcal{E}_L(t)$, at the time,~$t_0$, the electron is released into the continuum, $e$ is the charge of an electron and $\vec{p}_0$ is the momentum of the electron in the absence of the IR field. 
In this semi-classical approximation, the interaction with the laser maps the arrival time of the x-ray pulse to the angle of the electron momentum shift~(or streaking angle).
In the presence of a short-range potential, the interaction between the outgoing electron and the potential modifies the streaking angle, and the apparent emission time of the electron is effectively delayed with respect to the arrival time of the ionizing pulse.
This delay can be attributed to the group velocity of the outgoing electronic wavepacket~\cite{schultze_delay_2010}, as shown in the supporting material, and can be related to the kinetic energy dispersion of the phase of the photoionization matrix elements~\cite{pazourek_attosecond_2015}.

To define a delay, we must specify a reference event. 
We measure the photoemission delay of low energy electrons from the oxygen $K$-shell of nitric oxide with respect to $>120$~eV electrons emitted from the nitrogen $K$-shell of the same molecule ionzied by the same attosecond x-ray pulse.
The photoemission delay of the high-energy nitrogen $K$-shell electrons is not strongly affected by correlation effects and can be calculated with confidence to be less than $\sim5$~as with reference to the arrival of the x-ray pulse~\cite{dahlstrom_theory_2013,serov_interpretation_2013, pazourek_attosecond_2015}.
The photon energy was scanned in $0.5$~eV steps across the oxygen $K$-edge near-edge X-ray absorption fine structure~(NEXAFS) region~($540-580$~eV) while measuring the momentum shift experienced by the oxygen and nitrogen $K$-shell electrons.

\begin{figure}
    \centering
    \includegraphics[width=1\textwidth]{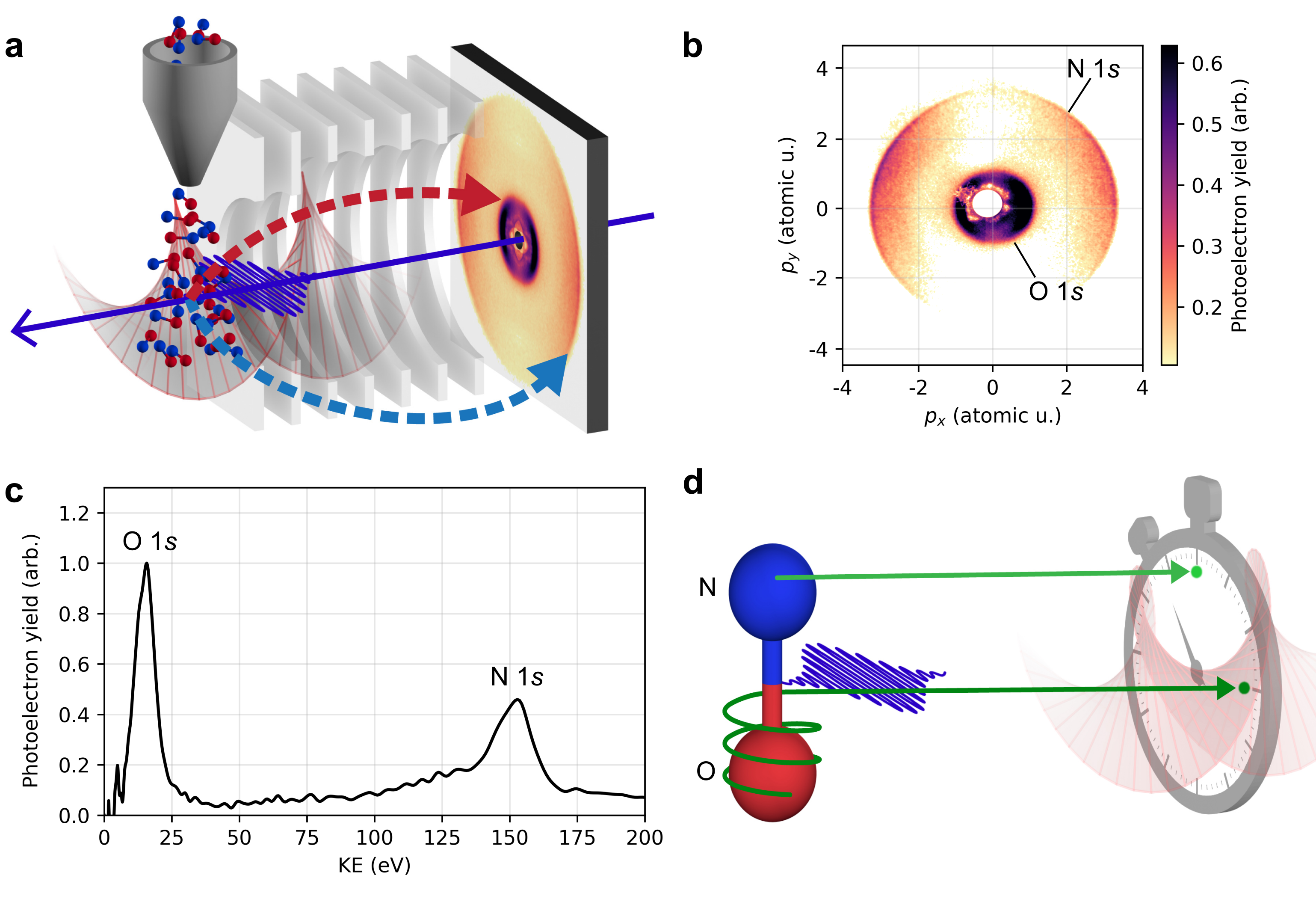}
    \caption{\textbf{a} Experimental setup showing electrons ionized by the x-ray pulse~(purple) from the nitrogen~(blue) and oxygen~(red) $K$-shells incident on the detector of the cVMI~\cite{li_co-axial_2018} spectrometer. \textbf{b} Momentum distribution of photoelectrons in the absence of the streaking field at an x-ray central photon energy of 563 eV. \textbf{c} Kinetic energy distribution by inverse Abel transformation of momentum distribution in panel \textbf{b}. \textbf{d} Electrons ionized from the nitrogen and oxygen $K$-shells, respectively, experience different trajectories in the molecular potential, resulting in a relative photoemission delay. (see text for details)}
    \label{fig:expt_setup}
\end{figure}

Our measurement was performed at the Atomic, Molecular and Optical physics~(AMO) experimental hutch at the Linac Coherent Light Source~(LCLS), using attosecond x-ray pulses produced by enhanced self-amplified spontaneous emission~(ESASE)~\cite{zholents_method_2005,duris_tunable_2020,li_attosecond_2022}.
Photoelectrons produced by X-ray ionization were streaked by combining
attosecond x-ray pulses with a co-propagating $2.3$~$\mu$m wavelength, circularly polarized laser pulse with $\sim100$~fs FWHM duration.
The momentum distribution of the emitted photoelectrons was recorded using a co-axial velocity map imaging (c-VMI) spectrometer, as shown in Fig.~\ref{fig:expt_setup}, which was designed for measurement of high energy electrons~\cite{li_co-axial_2018,li_attosecond_2022}.

The vector potential of the circularly polarized streaking pulse rotates with angular velocity $\nicefrac{2\pi}{T}$, where $T_L=7.7$~fs for a wavelength of $2.3~\mu$m.
Therefore the difference in momentum shift $-\vec{A}(t_i)$ between different photoelectron features produced by the same x-ray pulse encodes the delay between the two photoemission events~\cite{li_attosecond_2022}.
Because the period $T_L$ of the circularly polarized laser was well-known in our experiment, if the momentum shift between two photoelectron features is $\Delta\theta$, the photoemission events were separated by a time $\Delta\tau$:
\begin{linenomath}\begin{equation}
\Delta\tau=\frac{\Delta\theta}{2\pi}\times T_L.
\label{eqn:ang_to_time}
\end{equation}\end{linenomath}

\begin{figure}
    \centering
    \includegraphics[width=\textwidth]{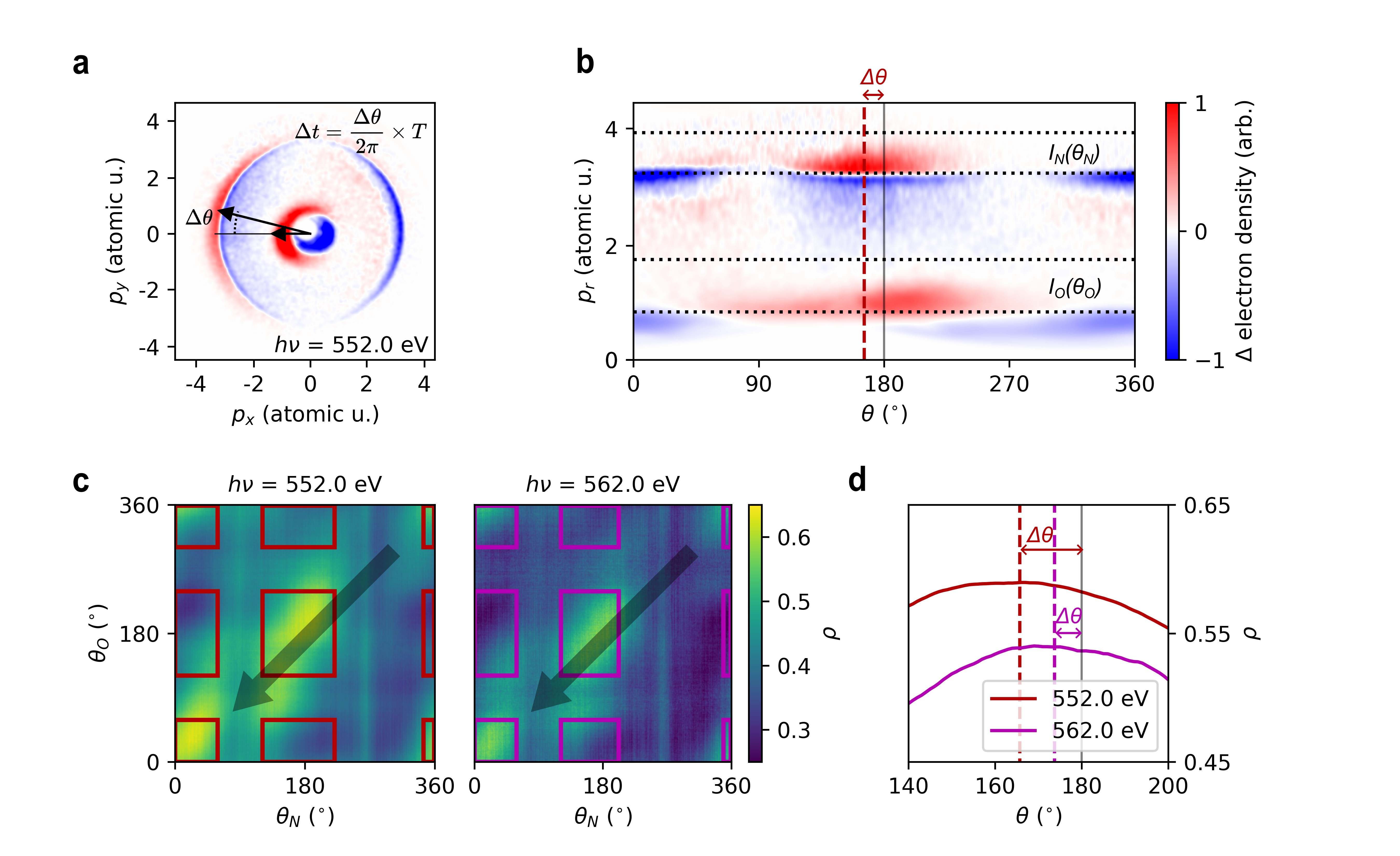}
    \caption{\textbf{a} Differential photoelectron momentum distribution~(see text) induced by the IR streaking laser for only one x-ray/laser arrival time. The relative photoemission delay is proportional to the difference in angle $\Delta \theta$ between the momentum shift of the oxygen and nitrogen photolines. This distribution is rebinned to polar coordinates in panel \textbf{b}, and the vectors $I_N(\theta_N)$ and $I_O(\theta_O)$ are defined by integrating the labeled regions along the radial coordinate. \textbf{c} Partial correlation maps recorded at two different photon energies. The marked regions of the maps are averaged in the direction indicated by the black arrows to produce the traces shown in panel \textbf{d}. The offset of the position of maximum correlation gives the angular difference in momentum shift between the two photoelectron distributions.
    }
    \label{fig:data_analysis}
\end{figure}

The momentum shift can be seen in the data plotted in Figure~\ref{fig:data_analysis}, which shows the differential electron momentum distribution measured for $552$~eV x-rays in Cartesian~(a) and polar~(b) coordinates.
The electron momentum distribution is plotted as a difference image between measurements where the direction of vector potential of the IR laser at the time when the x-ray pulse ionizied the sample made a $170^{\circ}$ angle with respect to the x-ray polarization~(the long arrow in panel~(a) and the dashed-red line in panel(b)) and measurements where the IR laser was intentionally mistimed with respect to the x-ray pulse. 
The momentum shift of the electrons ionized from the nitrogen $K$-shell (high energy feature) is in the direction of the IR vector potential at the time of ionization. 
It is clear for panels~(a)~and~(b) that the momentum shift of the lower energy, oxygen $K$-shell ionization feature, is different, implying a delay in the emission of the oxygen $K$-shell electrons. 

\begin{figure}
    \centering
    \includegraphics[width=0.65\textwidth]{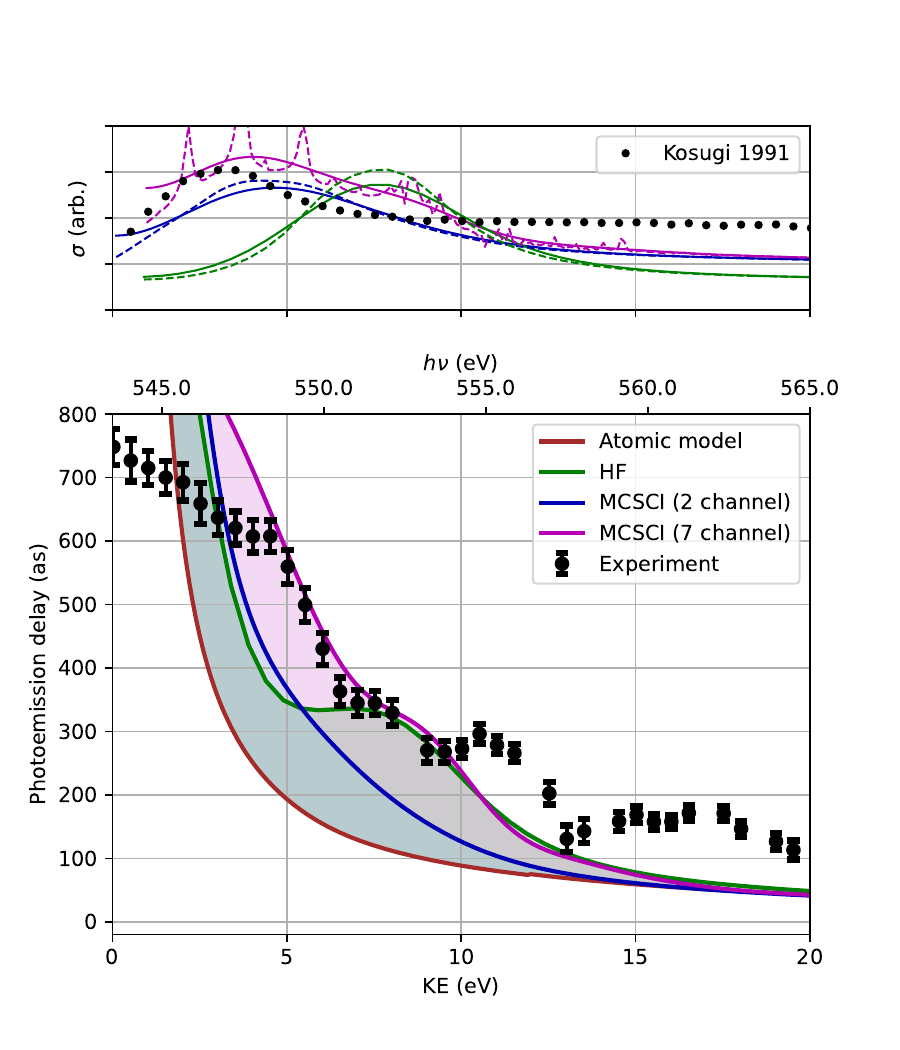}
    \caption{
        Oxygen $K$-shell photoionization cross-section~(upper panel) in the Hartree Fock approximation~(green), two-channel~(blue), and seven-channel~(magenta) Schwinger configuration interaction method. The calculated result is shown as a dashed curve, and the solid curve shows the results convoluted with the expected photon energy distribution from the XFEL.   
        The black dashed line shows the photoabsorption cross-section measured by Kosugi~\textit{et al.}~\cite{kosugi_highresolution_1992}.
        The experimental photoemission delay~(black points) is shown in the lower panel along with calculated photoemission delay.  
    }
    \label{fig:exp_dels}
\end{figure}

\paragraph{Data Analysis}
To measure the differential direction of the momentum shift between the high energy electrons ionized from the nitrogen $K$-shell~(N$1s$) and the low energy electrons ionized from the oxygen $K$-shell~(O$1s$) and extract the relative photoemission delay using Eqn.~\ref{eqn:ang_to_time}, we exploit the inherent temporal jitter between the x-ray pulse and the circularly polarized laser~($\sim500$~fs full-width-at-half-maximum~(FWHM)~\cite{glownia_time-resolved_2010}).
The scheme is illustrated in the lower panels of Fig.~\ref{fig:data_analysis}.
%
The temporal jitter between the x-rays and the streaking field results in shot-to-shot variation the in momentum shift of the oxygen and nitrogen $K$-shell photoemission features, but the difference between these two angles remains fixed.
This difference can be isolated in the partial correlation map~(defined in the methods section) between the angle-resolved electron yield in two different regions of our detector.
These regions correspond to the high kinetic energy side of the N$1s$ and O$1s$ photoemission features respectively, as shown in panel~\textbf{b} of Fig.~\ref{fig:data_analysis}.
Calculating the partial correlation removes the spurious correlation due to shot-to-shot fluctuations in the FEL, pulse energy, spectrum, etc.
The angular difference between the momentum shift of the two features is encoded in the offset between the peak of the partial correlation coefficient from the diagonal of this map, shown in panel~\textbf{c} of Fig.~\ref{fig:data_analysis}.
This analysis is further described in the Supplementary Information~(SI).
The measured kinetic energy-dependent photoemission delay between the O$1s$ and N$1s$ photoelectrons is shown in Fig.~\ref{fig:exp_dels}.

The photoelectron spectrum shown in Fig.~\ref{fig:expt_setup}~(c) contains two features, one corresponding to an ionization which leaves the residual cation in a $^3\Sigma$ state and another where the residual cation is found to be in a $^1\Sigma$ state.
The binding energy difference between these two features is $\sim 0.7$~eV~($543.3$~eV$/544$~eV, respectively~\cite{kosugi_highresolution_1992}) and is not resolved in this experiment.
Thus the experimental data presented in Fig.~\ref{fig:exp_dels} is a mixture of both ionization channels taking into account that the $^3\Sigma$-channel has a three-fold higher cross-section.  

\paragraph{Discussion}
The measured photoemission delay of the low energy electrons from the oxygen $K$-shell~(black points in Fig.~\ref{fig:exp_dels}) decreases with increasing kinetic energy, as expected.
The group delay incurred by a photoelectron wavepacket in a simple Coulomb potential is given by the derivative of the argument of the Coulomb phase-shift, $\partial_{E}\sigma_{l}$, where, $\sigma_{l}=\arg[\Gamma(l+1-\frac{iZ}{k})]$, $l$ is the angular momentum of the photoelectron, $Z$ is the charge of the residual ion and $\nicefrac{k^2}{2}$ is the kinetic energy of the photoelectron.
This simple analytical form~(plotted in brown in Fig. \ref{fig:exp_dels}) accounts for the expected photoemission delay from a centrosymmetric system, such as an atom.
There is an additional contribution to the extracted photoemission delay due to fact that the photoelectron is being driven by the streaking laser field in the presence of the long-range Coulomb potential of the ion.
This so-called Coulomb laser coupling~(CLC) delay was calculated for the case of circular polarization using the classical trajectory Monte Carlo~(CTMC) method, which is described further in the SI, and then subtracted from the pure Coulomb delay in Fig.~\ref{fig:exp_dels}.
These purely Coulombic effects describe the general trend of the measured photoemission delay, i.e. a decreasing delay with increasing photoelectron kinetic energy, however the absolute agreement is poor, pointing to the importance of molecular structure in the attosecond delay dispersion.

To understand our measured delay, we compare with numerical calculations of photoionization of the oxygen $K$-shell of nitric oxide.
We initially employ a single center expansion~(SCE) for the continuum wavefunction, solving a system of Hartree-Fock~(HF) equations~\cite{mountney_streaking_2023}.
The bound states of the core-excited NO$^{+*}$ ion are computed using the HF method with the \textit{aug-cc-pVQZ} basis set.
We choose to use the bound states of the core-excited NO$^{+*}$ ion to better approximate the orbital contraction that occurs during ionization. 
Using these states we calculate the photoionization matrix element~(PME) in the dipole approximation. 
From the PME, we can calculate the orientation- and electron emission direction-resolved cross-section and delay, as described in the methods~\cite{baykusheva_theory_2017}. 
The calculation of the photoemission delay is then convolved with a gaussian kernel of width $\sigma=1.3$~eV to account for the inherent spectral jitter in the XFEL photon energy~\cite{duris_tunable_2020, li_attosecond_2022}.
The green curves in Fig.~\ref{fig:exp_dels} show the calculated delay~(and cross-section in the upper panel).
At the HF level of electronic structure theory, the calculated photoemission delay~(green curve) is much closer to the measured delay, compared to the atomic theory~(brown curve). 

For a molecular system, the photoemission delay maps to molecular effects arising from the complex shape of the potential of the residual molecular ion \cite{haessler_phase-resolved_2009,huppert_attosecond_2016, biswas_probing_2020}.
In the case of $K$-shell ionization of nitric oxide, there is noted increase in the photoionization cross-section $\sim4$~eV above the $K$-shell threshold, as shown by the measurements of Kosugi~\textit{et al.}~\cite{kosugi_highresolution_1992} reproduced in the upper panel of Fig.~\ref{fig:exp_dels}.
This increase has been postulated to result from the existence of a shape resonance in the electronic continuum. 
The shape resonance occurs when the outgoing photoelectron experiences a transient trapping due to a combination of the molecular potential and the centrifugal barrier created by the photoelectron angular momentum~\cite{piancastelli_neverending_1999}.
The electron can tunnel through the barrier and leave the molecular potential, resulting in an increase in the photoemission time delay.

The HF calculation shows a similar increase in the photoionization cross-section above the $K$-shell threshold, although the energetic position of the maximum in the cross-section is shifted by several electron volts compared to the measurement. 
Turning to the calculated photoemission delay, we can see that this increase in cross-section coincides with an increase in the calculated photoemission delay, which is indicative of a molecular shape resonance. 
  
It is worth noting that the shape resonance only appears in the HF calculation when we use the core-excited NO$^{+*}$ bound states in the HF calculation the continuum wavefunction~\cite{mountney_streaking_2023,li_photoemission_2007}.
Using the bound states from the neutral NO molecule fails to produce a shape resonance in the continuum.  
In contrast the valence ionization, which creates a delocalized hole in the electronic density of the molecule, ionization of a highly-localized core-level causes the electron density in the ion to contract around the core-level vacancy, a consequence of the highly correlated nature of the core-level electrons~\cite{breidbach_universal_2005, lin_theoretical_2001, li_photoemission_2007}. 
By using pre-contracted orbitals in the calculation we can approximate the correlation interaction of the core-level electrons even though we perform a mean field calculation. 

While the HF calculation is able to explain some of the structure in the photoemission delay, the theory fails to reproduce the entire measurement.
In an attempt to better describe the delays we calculate the photoionization matrix element using the multichannel Schwinger configuration interaction~(MCSCI) method.
The orientation- and electron emission direction-resolved cross-section and delay for two channels~($^1\Sigma$ and $^3\Sigma$) and seven channels~($^1\Sigma$, $^3\Sigma$, and five additional shake-up states) are shown as the blue and purple curves in Fig.~\ref{fig:exp_dels}, respectively. 
Again these results have been convolved with the same gaussian kernel to account for the jitter in the XFEL photon energy.
Both the two-channel and seven-channel calculations predict a shape resonance just above the $K$-shell threshold, which results in an increase in the photoionization cross-section consistent with the previous measurements. 
However, the expected increase in the photoionziation delay due to the presence of the shape resonance is less pronounced in the two-channel calculation~(compared to the HF calculation).
This is a result of the shape resonance shifting to lower kinetic energy, where the Coulomb contribution to the photoemission delay is much larger.  
Thus, the two-channel calculation provides an improvement over the HF calculation in the description of the photoionizaiton cross-section, but does not significantly alter the agreement between the measured and calculated photoemission time delay. 

Turning our attention to the seven-channel model, the photoionization cross-section~(dashed magneta line) in the top panel of Fig.~\ref{fig:exp_dels} shows a series of sharp autoionizing shake-up features, in addition to a shape resonance.
While these features appear quite distinct, once the calculation is averaged over several N-O bond lengths (which shifts their energetic position, see SM) and convolved with the expected bandwidth and energy jitter of the XFEL, the cross-section again appears smooth~(solid curves in upper panel of Fig.~\ref{fig:exp_dels}).
There is however a marked effect on the photoemission delay, which increases significantly between the two-channel and seven-channel calculation~(purple shaded area in Fig.~\ref{fig:exp_dels}). 
The seven-channel calculation greatly improves the agreement with the measured photoemission delay.  
It is striking that the x-ray photoemission delay is strongly dependent on channel-coupling in the final ionic state, demonstrating the sensitivity of x-ray photoemission delay for probing electron correlation effects.
Our measurements confirm that the increase in photoioniztion cross-section just above the oxygen $K$-shell threshold is due to the combination of resonant features in the continuum, both a shape resonance, and unresolved autoionizing states. 

For electrons with less than $\sim3$~eV kinetic energy, we expect the separability of the photoionization delay and the CLC contribution to break down.
The disagreement between our measurement and the calculation at higher electron kinetic energies is likely the result of additional ionization channels, not considered in the MCSCI calculation. 
For example, in a previous measurement of the symmetry-resolved ionization cross-section, Kosugi~\textit{et al.}~\cite{kosugi_highresolution_1992} observed an increase in the ionization asymmetry parameter near an electron kinetic energy of $\sim17$~eV, possibly indicating additional structure in the ionization continuum. 

\begin{figure}
    \centering
    \includegraphics[width=1\textwidth]{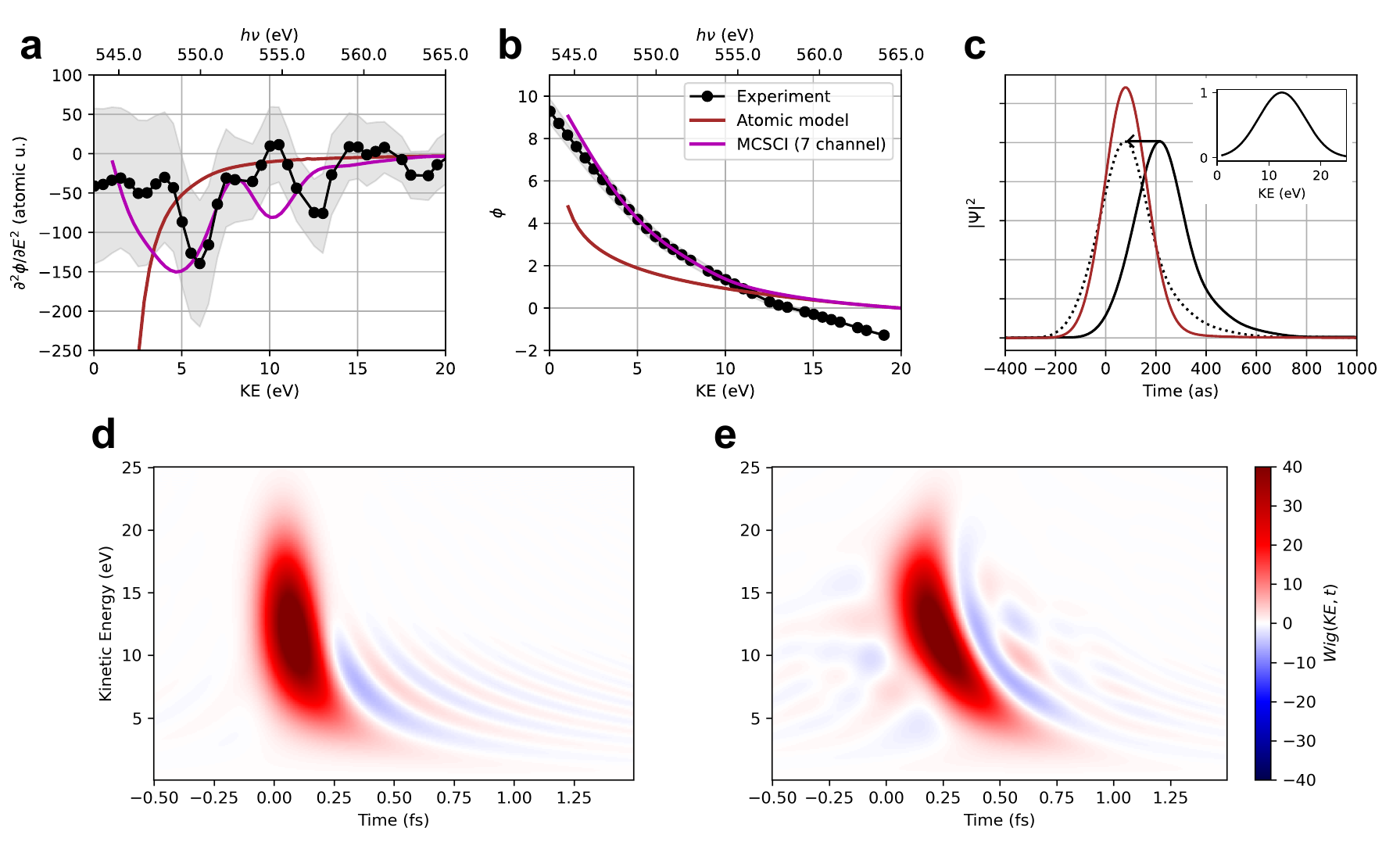}
    \caption{First derivative with respect to energy (a) and integral (b) of the photoemission delays from Fig.~\ref{fig:exp_dels}(b). The black curve is derived from the measured data and the shaded gray area shows the standard error. The brown curve shows the result calculated for an atomic potential and the magenta curve shows the result from the 7-channel MCSCI calculation. The reshaping of the photoionizied wavepacket due to the molecular potential is shown in panels (c)-(e). Panel~(d) shows the Wigner quasi-probability distribution of the photoionizied wavepacket in the atomic potential, and panel~(e) shows the wavepacket reconstructed from the experimental phase shown in panel~(b). The marginals~(or projection) of the Wigner distribution along the energy axis are shown in  panel~(c):The brown curve in the main panel shows the wavepacket from an atomic potential. The black curve shows the result from the experimental phase. To illustrate the reshaping the dashed-black curve is the experimental curve shifted to overlap with the centroid of the atomic wavepacket. The inset in panel~(c) shows the marginal along the time-axis, which shows the spectral content of the electronic wavepacket.
    }
    \label{fig:dels_derive}
\end{figure}

By employing a fully tunable attosecond XFEL, we have mapped out the photoemission delay in $0.5$~eV steps between the $K$-shell ionization threshold~($543$~eV) and $565$~eV.
This is unique in photoemission delay measurements, which are usually performed at a fixed photon energy.
This enhanced fidelity allows us to make better comparison with theory, as we can derive other observables from the measurement. 
For instance, Fig.~\ref{fig:dels_derive}~(a) shows the first derivative of the photoemission delay. 
If the photoemssion delay represents some orientation and emission-angle average of the group-delay of the photoelectron wavepacket~\cite{schultze_delay_2010}, then the quantity shown in this panel represents the group-delay dispersion~(GDD), or chirp, of the average wavepacket.
For electron kinetic energies above $5$~eV, the 7-channel MCSCI results shows good agreement with the measurement. 
There is a notable exception above $10$~eV, which suggests there are additional ionization channels which are not captured in the theory.  

Fig.~\ref{fig:dels_derive}~(b) shows the integral of the photoemission delay. 
This can be directly compared to the orientation- and emission-angle-averaged phase of the dipole matrix element. 
Again the 7-channel MCSCI result shows excellent agreement with the measurement. 
We can use this phase to approximate the orientation- and emission-angle-averaged electronic wavepacket.
This wavepacket can be visualized with the Wigner quasi-probability distribution shown in Fig.~\ref{fig:dels_derive}~(e), which is a joint time-frequency representation of the photoionizied electron.
The marginal~(or projection) of the Wigner distribution along the energy-axis, shown in Fig.~\ref{fig:dels_derive}~(c), gives the time-dependent probability~($|\psi(t)|^2$, black). 
The spectral amplitude of the wavepacket is taken from the time-axis marginal and is shown as an inset in panel~(c). 
We compare to the case of an atomic reference potential~(Figure~\ref{fig:dels_derive}~(d) and brown curve in panel~(c)).
The molecular potential creates an asymmetry in the electronic wavepacket with a sharp rising edge and an extended tail. 
The Wigner distribution shows us that this extended tail comes from the low energy electrons which are significantly delayed by the molecular potential~(in contrast to an atomic reference). 
The Wigner distribution has additional structure near $7$~eV and $13$~eV, where the quasi-probability is extended along the time-axis. 
From the MCSCI theory we have identified the $7$~eV structure as originating from the combination of a shape resonance and additional autoionizing structures. 
We postulate that the temporally-elongated feature at $13$~eV likely originates from autoionization $via$ an additional channels not considered in our current model.  
The additional analysis, afforded by the density of measurements, provides greater discernment of the theoretical predictions, and offers deeper insight into the scattering and re-shaping of the outgoing electron due to the molecular potential.

There is an additional consideration in $K$-shell photoionization close to the ionization threshold. 
Unlike in valence ionization, the core-ionized system produced by x-ray photoionization is highly unstable and (in the case of low-Z systems) will decay within several femtoseconds $via$ the process of Auger-Meitner decay.
The core vacancy is filled by a valence electron, and the resultant excess energy in the system is released as a high kinetic energy electron.
When the $K$-shell photoelectron is very slow it will be overtaken by the Auger-Meitner electron, which results in a change in the effective ionic potential in which that photoelectron is propagating.
The loss in screening results in an additional delay for the photoelectron, which we account for using the classical model of Russek and Mehlhorn \cite{russek_post-collision_1986}.
We find that a complete description of x-ray photoemission delays also requires accounting for post-collision interaction following the ultrafast Auger-Meitner decay in the core-ionized system.
The method to account for this post-collision interaction is described in the SI, and the appropriate correction has been made to the curves shown in Fig.~\ref{fig:exp_dels}.

\paragraph{Conclusion}
In this work, we have demonstrated the measurement of attosecond photoemission delays at x-ray wavelengths.
This measurement was achieved using the experimental technique of attosecond angular streaking to probe the photoemission delay from different core-level orbitals in a molecular system.
We used attosecond x-ray pulses from an x-ray free electron laser, which has continuous wavelength tunability across the entire soft x-ray regime.
This unique feature of an attosecond XFEL allows us to probe the photoemission delay across the entire NEXAFS spectral region. 
X-ray induced photoelectron emission delays uniquely map the electronic environment of a molecule.
We measure an increase in the photoemission delay that we assign to the transient trapping of an ionized electron by the molecular potential very close to the oxygen $K$-shell threshold, a so-called shape resonance. 
We also find that the x-ray photoemission delay is highly sensitive to electron correlation effects: our theoretical description requires the inclusion of orbital relaxation and channel coupling in the final ionized state to explain the sizeable delay in the measurement.
In addition, the post-collision interaction between the outgoing photoelectron and the subsequent Auger-Meitner electron from the decay of the core-ionized state induces an additional increase to the photoemission delay measurement.
This effect has yet to be observed in photoemission studies, because only core-level ionization produces fast AM electrons. 
Our work provides a new experimental probe of electron correlation effects in quantum systems; by directly accessing the multi-electron dynamics of a fundamental process, the x-ray photoelectric effect, on its natural attosecond timescale.
Providing detailed understanding of electrons in the attosecond regime allows the determination of the properties of matter.

\paragraph{Acknowledgements:}
Use of the Linac Coherent Light Source (LCLS), SLAC National Accelerator Laboratory, is supported by the U.S. Department of Energy~(DOE), Office of Science, Office of Basic Energy Sciences~(BES) under Contract No. DE-AC02-76SF00515. 
This work was supported primarily by the AMO Physics Program, Chemical Sciences, Geosciences and Biosciences Division~(CSGB), BES, DOE. 
A.M. J.D., and Z.G. acknowledge support from the Accelerator and Detector Research Program of the Department of Energy, Basic Energy Sciences division. 
A.L. and L.O. were supported by DOE Investigator-Initiated Research grant, Award ID DE-SC0022093, of CSGB, BES, DOE.
P.R. and M.F.K. acknowledge support by the German Research Foundation through LMUexcellent. 
I.I. acknowledges support from the Institute for Basic Science grant (IBS-R012-D1).

\paragraph{Data Availability}
The data supporting the findings of this study are available from the corresponding authors upon request.

\section*{Methods:}

\paragraph{Generating sub-femtosecond X-ray pulses:}
The attosecond x-ray pulses were produced using the method described in Ref.~\cite{duris_tunable_2020}.

\paragraph{Streaking laser setup:}
X-rays generated in the undulators are focused with a pair of Kirkpatrick–Baez mirrors to a spot size of $\sim55~\mu$m diameter~(FWHM). 
The streaking laser pulse is generated using an optical parametric amplifier (TOPAS-HE, Light Conversion) pumped by the titanium-doped sapphire laser system~($10$~mJ, $\sim40$~fs, $800$~nm, $120$~Hz) synchronized to the accelerator.
The OPA is tuned to 2400 nm with a pulse energy of $200~\mu$J. 
A quarter-wave plate~(Thorlabs AQWP05M-1600) is used to produce circularly polarized laser pulses, which are then focused with a $750$~mm focal length CaF$_2$ lens. 
The streaking laser field is combined with the XFEL beam using a silver mirror with a 2-mm-diameter drilled hole, and both pulses come to a common focus in the interaction region of a coaxial velocity map imaging apparatus~\cite{li_co-axial_2018}. 
The laser is focused to a diameter of $\sim 110~\mu$m.

\paragraph{Partial Covariance Analysis:}
We calculate the partial correlation maps shown in Fig.~\ref{fig:data_analysis}~\textbf{b} by calculating the partial correlation coefficient between the radially integrated intensities $I_N$ and $I_O$ at each pair of angles $\theta_N, \theta_O$, according to~\cite{Johnson_Wichern_2007}: 
\begin{linenomath}\begin{equation}
\rho(\theta_N, \theta_O)=\frac{\PCov\left(I_N(\theta_N),I_O(\theta_O);P\right)}{ \sqrt{\PCov\left(I_N(\theta_N),I_N(\theta_N);P\right)\PCov\left(I_O(\theta_O),I_O(\theta_O);P\right)}},
\label{eq:sm:pcorr}
\end{equation}\end{linenomath}
where
\begin{align}
\PCov\left(X,Y;I\right) = \Cov\left(X,Y\right) - \frac{\Cov\left(X,I\right)\Cov\left(I,Y\right)}{\mathrm{Var}\left(I\right)},
\end{align}
and
\begin{align}
\Cov\left(X,Y\right) = \langle XY \rangle - \langle X \rangle \langle Y \rangle.
\end{align}
The angular brackets denote the average across multiple XFEL shots.
The partial correlation controls for the single-shot x-ray pulse intensity~($P$), which is an inherently fluctuating value for SASE FELs.
From the partial correlation maps, we identify the offset of the maximum from the diagonal line described by $\theta_N = \theta_O$, in those regions of the detector where the radial gradient of the measured momentum distribution dominates, i.e. the highlighted regions in Fig.~\ref{fig:data_analysis}~(c). 
As described in section~4 of the supporting material, this offset corresponds to the angular difference between the momentum shift of the two photoemission features.
We identify the offset by transforming the axes to ($\theta_O, \pi + \theta_O - \theta_N$) and averaging across $\theta_N$ to produce a one-dimensional trace.
The peak of this trace corresponds to the difference in the angular direction of the momentum shift imparted on the two photolines by the streaking laser.
We identify the peak by finding the root of the derivative of the polynomial which fits this trace $\pm 40~{^\circ}$ about the maximum. 
The quoted error bar is the uncertainty in this root due to the uncertainty in the fit coefficients for the polynomial.

The method of modeling of the photoionization matrix element within the Hartree Fock framework using the aug-cc-pVQZ basis set \cite{dunning_gaussian_1989} is described in detail in references \cite{banks_interaction_2017} and \cite{mountney_streaking_2023}.
The MCSCI calculations are performed according to the method described in \cite{li_photoemission_2007}.
Further information on these calculations is provided in the supporting material. 
The molecular-frame photoionization dipole moment is expanded is a spherical harmonic basis:
\begin{linenomath}\begin{equation}
    \left\langle \psi^{(-)}_{i,\vec{k}}|\vec{r}|\Psi_0\right\rangle = \sum_{l,m\mu} I^{(i)}_{lm\mu}(k) Y_{l,m}(\hat{k})\mathcal{D}^{(1)}_{\mu0}(\hat{R}),
\end{equation}\end{linenomath}
where the subscript $i$ refers to the photoionization channel, $\vec{k}$ is the momentum of the photoionizied electron,  $\Psi_0$ is the ground state of the molecular system, $Y_{l,m}$ is the spherical harmonic, and $\mathcal{D}^{(l)}_{m_1,m_2}$ is the Wigner-D matrix which rotates the laboratory frame into the molecular frame, $\hat{R}$.
The photoemission delay is calculated from $I^{(i)}_{lm\mu}(k)$ according to~\cite{baykusheva_theory_2017}: 
\begin{linenomath}\begin{equation}
     \tau=\int d\hat{k} \int d{\hat{R}}~\frac{|\sum_{l,m}I_{\hat{R},l,m}Y_{l,m}(\hat{k})|^2}{\sum_{l,m}|I_{\hat{R},l,m}|^2}\times \\ \frac{\partial}{\partial E} (\sum_{l,m}I_{\hat{R},l,m}Y_{l,m}(\hat{k})),
     ~\label{eqn:sm:dly}
\end{equation}\end{linenomath}
where the orientation- and emission-angle-delay is averaged over over all molecular orientations $\vec{R}$ and all outgoing electron directions $\vec{k}$ weighted by the relative cross-section. 

\bibliography{referencesJPC.bib, references-additional.bib}
\end{document}


\maketitle

\section{Equivalence of the phase of the photoionization matrix element and a delay measured in angular streaking}
\label{sec:sm:phase-vs-dly}
Below, we develop a description of the effect of the phase of the dipole matrix element on the photoelectron distribution which is measured in an angular streaking experiment.
Our analysis is performed within the strong-field approximation~(SFA) of the laser matter interaction~\cite{kitzler_quantum_2002}:
\begin{equation}
     b(\vec{p},\tau) = i \int_{t_0}^{\infty} dt \vec{E}_X(t-\tau) \cdot \vec{d}\left(\vec{p}-\vec{A}(t)\right) e^{-i\Phi_V(t)},
     ~\label{eqn:sm:sfa1}
\end{equation}
where $|b(\vec{p},\tau)|^2$ is the probability of measuring an electron with momentum $\vec{p}$ following the combined interaction of the target system with the electric field of the x-ray pulse~($\vec{E}_X$), delayed by $\tau$, with respect to the peak of the field of the infrared laser field, $\vec{E}_L=-\partial_t\vec{A}$.
$\vec{d}(\vec{p})=\langle \vec{p}|r|\Phi_0\rangle$ is the photoionization dipole matrix element describing the x-ray ionization of the ground state~($\Phi_0$) with ionization potential $I_p$.
$\Phi_V$ in Eq.~\ref{eqn:sm:sfa1} is the so-called Volkov phase and is given by,
\begin{equation}
    \Phi_V(t) = \int_t^{\infty}dt^{\prime}\frac{1}{2}\left[\vec{p}-\vec{A}(t^\prime)\right]^2 - I_p t^{\prime}.
    \label{eqn:sm:volkov}
\end{equation}
In angular streaking experiments, the streaking laser field is circularly-polarized, and can be written as,
\begin{equation}
    \vec{A}(t) = A_0(t)\left[ \hat{x} \cos(\omega t) + \hat{y} \sin(\omega t) \right],
\end{equation}
where $A_0(t)$ is the slowly-varying envelope of the pulse and $T=\nicefrac{2\pi}{\omega}$ is the laser period. 
Expanding the expression for the Volkov phase, Eq.~\ref{eqn:sm:volkov}, yields
\begin{eqnarray}
    \Phi_V(t) &=& \int_t^{\infty}\left(\frac{p^2}{2}- I_p\right)dt^{\prime} \\
    &-& A_0(t) \left[p_x \int_t^{T_x}dt{^\prime} \cos(\omega t^\prime) + p_y \int_t^{T_y} dt{^\prime} \sin(\omega t^\prime) \right],
    \label{eqn:sm:volkov_expansion}
\end{eqnarray}
where we have assumed that the streaking laser field envelope is approximately constant over a single period of the field.
In addition we assume $|\vec{p}|\gg|\vec{A}_0|$ and thus neglect any terms proportional to $|\vec{A}(t)|^2$.
We have defined a few special times~($T_x$ and $T_y$) where the electric field~($\vec{E}_{IR}(t)=\partial_{t}\vec{A}(t)$) is directed along the $x-$ or $y-$axis respectively:  \begin{eqnarray}
\vec{A}(T_x) = A_0(T_x) \hat{y} \nonumber \\
\vec{A}(T_y) = A_0(T_y) \hat{x},
\end{eqnarray}
and $T_x=T_y+\frac{T}{4}$.
In writing Eq~\ref{eqn:sm:volkov_expansion} we have used the fact that
\begin{eqnarray}
    \int_{T_x}^{\infty}dt{^\prime} A_0(t^\prime)\cos(\omega t^\prime) = \int_{T_y}^{\infty}dt{^\prime} A_0(t^\prime)\sin(\omega t^\prime) = 0.
\end{eqnarray}
The expansion in Eq.~\ref{eqn:sm:volkov_expansion} demonstrates the interesting property of the SFA; the momentum shift of the electron in a streaking experiment is acquired in the first quarter cycle of propagation, i.e. between $t$ and $(T_x,T_y)$.
Then we can rewrite Eq.~\ref{eqn:sm:sfa1} as,
\begin{eqnarray}
    b(\vec{p},\tau) &=& i \int_{t_0}^{\infty} dt \vec{E}_X(t-\tau) \cdot \vec{d}\left(\vec{p}-\vec{A}(t)\right) e^{-i\int_t^\infty(\frac{p^2}{2}-I_p)} \nonumber \\
    && \times \mbox{Exp}\left[iA_0(t)\left(p_x\int_t^{T_x}dt^\prime\cos(\omega t^\prime)+p_y\int_t^{T_y}dt^\prime\sin(\omega t^\prime)\right)\right].
    \label{eqn:sm:sfa2}
\end{eqnarray}

To show that the phase of the dipole matrix element can be related to a time delay of the x-ray electric field we consider the same exemplary case as Schultze~\textit{et al.}~\cite{schultze_delay_2010}, namely we define a photoionization matrix element with a phase that depends linearly on the electron energy:  
\begin{equation}
    \vec{d}(\vec{p})= \hat{d}\left|{d}(\vec{p})\right|e^{-i\alpha(\vec{p})\frac{p^2}{2}} = \frac{\vec{p}}{\left(\vec{p}^2+2I_p\right)^3} e^{-i\alpha(\vec{p})\frac{p^2}{2}}.   
    \label{eqn:sm:dipole}
\end{equation} 
In the second equality we have assumed that magnitude of the dipole matrix element is given by the first-order Born approximation to ionization of a hydrogenic $1s$ system.
We can write 
\begin{eqnarray}
\label{eqn:sm:dipole_expand}
\vec{d}\left(\vec{p}-\vec{A}(t)\right) & = & \left| {d}\left(\vec{p}-\vec{A}(t)\right) \right| e^{-i\alpha\frac{\left[\vec{p}-\vec{A}(t)\right]^2}{2}} \\
&\sim& \left| {d}\left(\vec{p}-\vec{A}(t)\right) \right| \mbox{Exp}\left\{-i\alpha\left[\frac{p^2}{2} - A_0(t)\left(p_x\cos(\omega t)+p_y\sin(\omega t)\right) \right]\right\} \nonumber \\
&=& \left| {d}\left(\vec{p}-\vec{A}(t)\right) \right| e^{-i\alpha\frac{p^2}{2}} \mbox{Exp}\left[i\alpha A_0(t)\left(p_x \int_t^{T_x}\omega \sin(\omega t^\prime)dt^\prime - p_y \int_t^{T_y}\omega \cos(\omega t^\prime)dt^\prime\right)\right]. \nonumber
\end{eqnarray}
Then plugging Eq.~\ref{eqn:sm:dipole_expand} into Eq.~\ref{eqn:sm:sfa2} we arrive at the expression,
\begin{eqnarray}
    b(\vec{p},\tau) &=& i \int_{t_0}^{\infty} dt \vec{E}_X(t-\tau) \left| {d}\left(\vec{p}-\vec{A}(t)\right) \right|
    e^{-i\left(\int_t^{\infty}\left(\frac{p^2}{2}- I_p\right)dt^{\prime} + \alpha\frac{p^2}{2}\right)} \nonumber \\
    &&\times \mbox{Exp}\left[i\sqrt{1+(\alpha\omega)^2}A_0(t)\left(p_x\int_t^{T_x}\cos(\omega [t + \tilde{\tau}_\alpha])+p_y\int_t^{T_y}\sin(\omega [t + \tilde{\tau}_\alpha])\right)\right] \nonumber \\
    &=& i \int_{t_0}^{\infty} dt \vec{E}_X(t-\tau) \left| {d}\left(\vec{p}-\vec{A}(t)\right) \right| e^{-i\int_{t-\tau_\alpha}^{\infty}\left(\frac{p^2}{2}- I_p\right)dt^{\prime}} \nonumber \\
    &&\times \mbox{Exp}\left[i\sqrt{1+(\alpha\omega)^2}A_0(t)\left(p_x\int_{t-\tilde{\tau}_\alpha}^{T_x}\cos(\omega t)+p_y\int_{t-\tilde{\tau}_\alpha}^{T_y}\sin(\omega t)\right)\right] \nonumber \\
    &\sim& i \int_{t_0}^{\infty} dt \vec{E}_X(t-\tau+\tau_\alpha) \left| {d}\left(\vec{p}-\vec{A}(t+\tau_\alpha)\right) \right| e^{-i\int_{t}^{\infty}\left(\frac{p^2}{2}- I_p\right)dt^{\prime}} \nonumber \\
     &&\times \mbox{Exp}\left[i\sqrt{1+(\alpha\omega)^2}A_0(t)\left(p_x\int_{t}^{T_x}\cos(\omega t)+p_y\int_{t}^{T_y}\sin(\omega t)\right)\right]
    \label{eqn:sm:sfa3}
\end{eqnarray}
where $\tilde{\tau}_\alpha=\frac{1}{\omega}\arctan(\alpha\omega)\sim\alpha=\tau_\alpha$, for small values of $\alpha\omega$, and we have used the identities, 
\begin{eqnarray}
    \alpha\omega \cos(\omega t) + \sin(\omega t) = \sqrt{1+(\alpha\omega)^2}\sin(\omega t + \arctan(\alpha\omega)) \nonumber \\
    \alpha\omega \sin(\omega t) + \cos(\omega t) = \sqrt{1+(\alpha\omega)^2}\cos(\omega t + \arctan(\alpha\omega)). \nonumber 
\end{eqnarray}    
Moreover, if $\alpha$ is small compared to the laser period~($\frac{2\pi}{\omega}$), then $\left| {d}\left(\vec{p}-\vec{A}(t+\tau_\alpha)\right) \right|\sim\left| {d}\left(\vec{p}-\vec{A}(t)\right) \right|$.
Comparing Eq.~\ref{eqn:sm:sfa3} to Eq.~\ref{eqn:sm:sfa2} reveals that the effect of the linear dependence of the dipole matrix element phase is to shift the x-ray field, $E_{X}$ along the time-axis by $\tau_\alpha$ and increase the apparent strength of the streaking field by $\sqrt{1+(\alpha\omega)^2}$.

\begin{figure}
    \centering
    \includegraphics[width=1\textwidth]{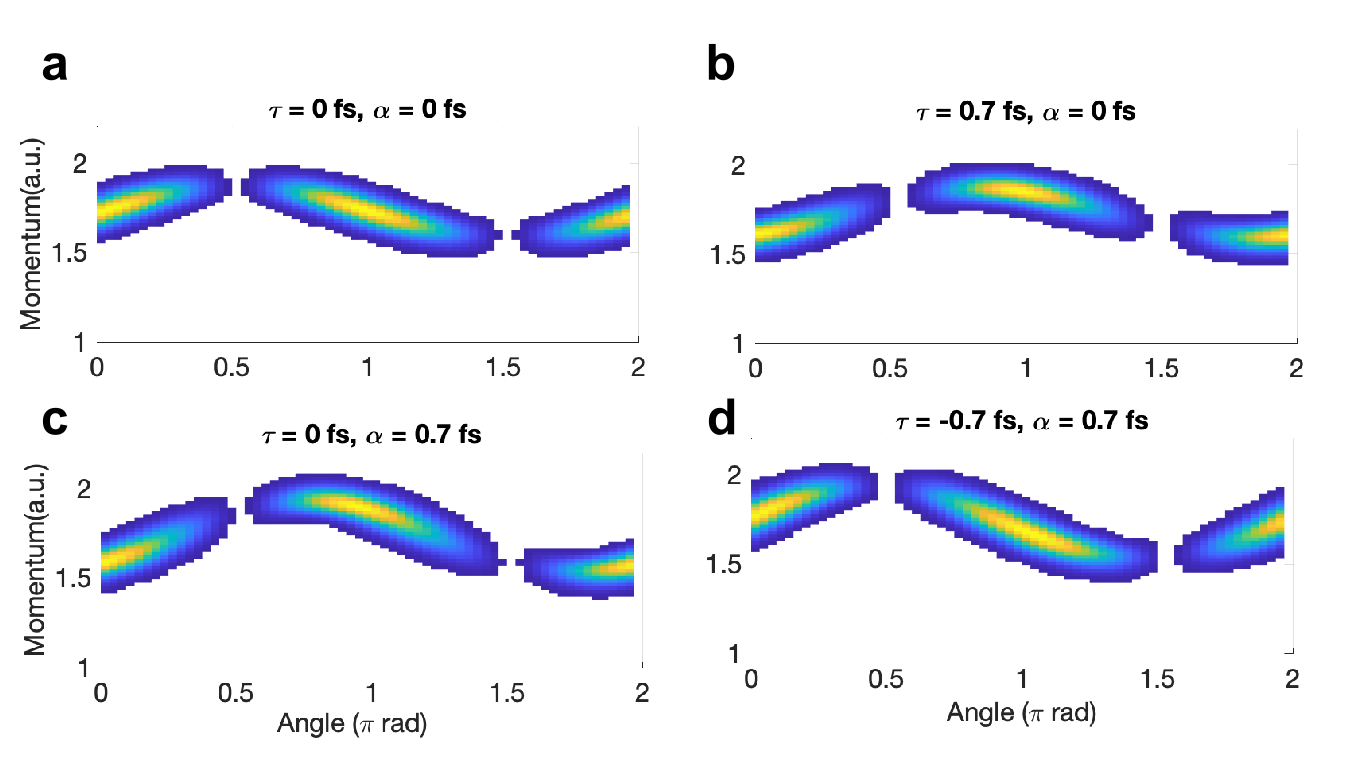}
    \caption{Simulated photoelectron momentum distribution in the strong field approximation~($|b(\vec{p},\tau)|^2$) for a $1.3~\mu$m~($T=6.7$~fs) laser field. Each panel shows the $p_z=0$ slice. Panel \textbf{a} shows the result for $\tau=0$~fs, $\alpha=0$~fs. In panel \textbf{b} the arrival time of the x-ray pulse is shifted by 700~as~($\tau=0.7$~fs, $\alpha=0$~fs). Panel \textbf{c} shows the result when a linear phase of $\alpha=0.7$~fs is applied ($\tau=0$~fs, $\alpha=0.7$~fs). Panel \textbf{d} shows the result when the linear phase is offset by the delay of the attosecond pulse~($\tau=-0.7$~fs, $\alpha=0.7$~fs).}
    \label{fig:sm:sfa_sim}
\end{figure}

We demonstrate the equivalence numerically by integrating Eq.~\ref{eqn:sm:sfa1} and comparing the final electron momentum distributions. 
For these simulation we consider a $300$~as~(full width at half maximum) x-ray pulse with a central frequency of $54.4$~eV and a $1.3~\mu$m wavelength laser field interacting with a hydrogen atom.
We analyze the change in the measured electron momentum distribution from adding a delay of $700$~as to the arrival time of the x-ray pulse and adding group delay of $\alpha=700$~as to the photoelectron wavepacket.
The calculated momentum distributions are shown in Fig.~\ref{fig:sm:sfa_sim}.

\section{Modeling of Photoemission Delay}
\label{sm:ModelPED}

\begin{figure}
    \centering
    \includegraphics[width=0.5\textwidth]{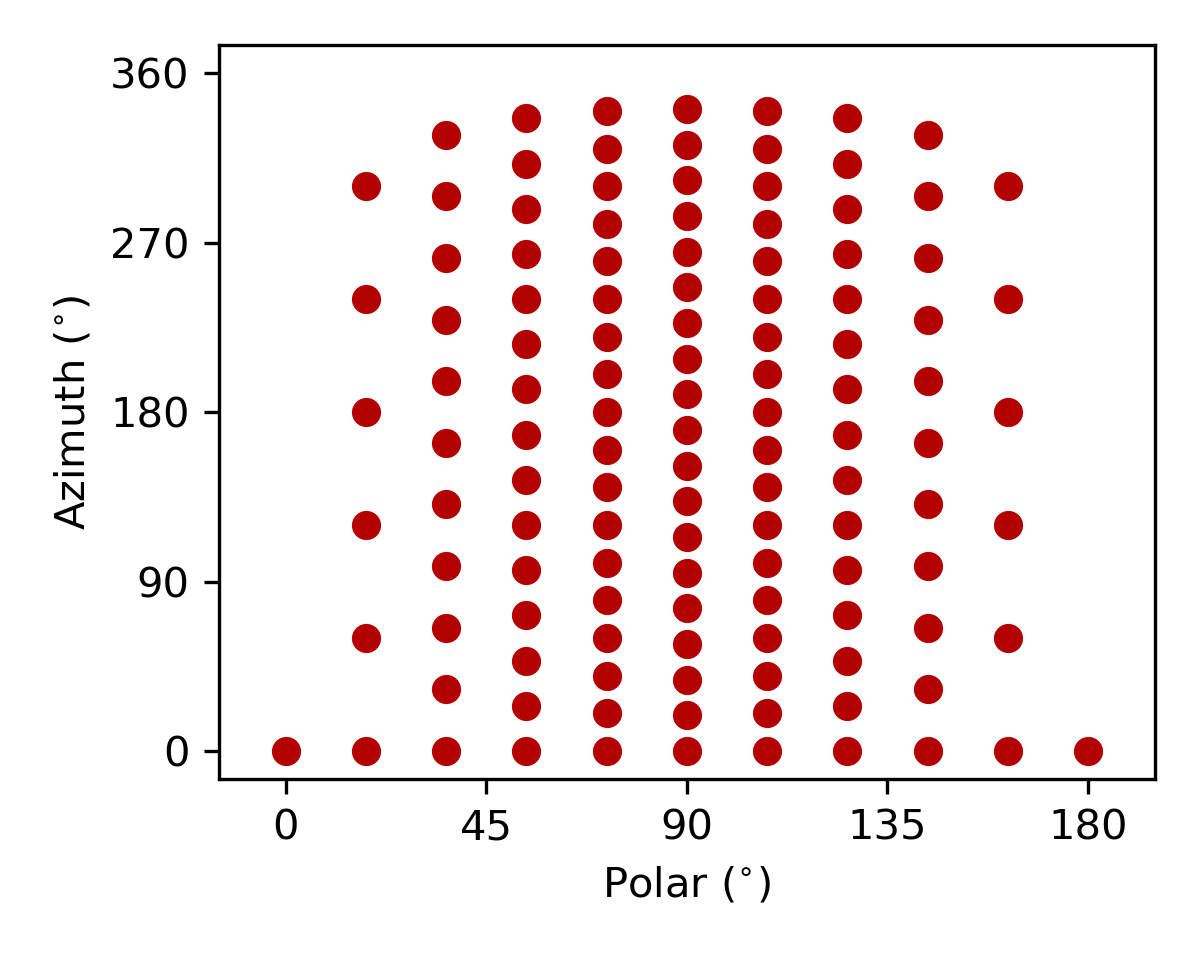}
    \caption{Distribution of polar and azimuthal angles used to define the orientation of the molecular axis with respect to the x-ray polarization axis.}
    \label{fig:sm:orientations}
\end{figure}

Our measurement is performed with randonly oriented molecular ensembles, meaning the calculations need to be summed over all orientations of the molecular axis with respect to the x-ray polarization axis.
In addition we need to sum over all out-going electron emission directions.
We evaluate the dipole matrix element for 121 orientations of the x-ray polarization with respect to the molecular axis, sampled uniformly on a sphere.
The distribution of polar and azimuthal angles describing these orientations is shown in Fig. \ref{fig:sm:orientations}.
The same set of angles is used to define 121 emission directions $\vec{k}$ of the outgoing photoelectron in the laboratory frame, and the (one-photon) photoemission delay $\tau$ weighted over all molecular orientations $\vec{R}$ and all outgoing electron directions $\vec{k}$ is found by weighting over cross-section according to \cite{baykusheva_theory_2017}:

\begin{equation}
     \tau=\int d\hat{k} \int d{\hat{R}}~\frac{|\sum_{l,m}I_{\hat{R},l,m}Y_{l,m}(\hat{k})|^2}{\sum_{l,m}|I_{\hat{R},l,m}|^2}\times \\ \frac{\partial}{\partial E} (\sum_{l,m}I_{\hat{R},l,m}Y_{l,m}(\hat{k})).
     ~\label{eqn:sm:dly}
\end{equation}

The ground state of the nitric oxide molecule is a $^2\Pi$ state with electronic configuration $1\sigma^2~2\sigma^2~3\sigma^2~4\sigma^2~1\pi^4~5\sigma^2~2\pi^1$.
An additional cross-section-weighted average of the photoemission delay is performed across orthogonal channels with different azimuthal quantum numbers of the initial ($M_i$) and final ($M_f$) molecular ion and channels for which the total (ion + electron) final state is a $\Sigma$ state with different (+/-) symmetry.

\begin{figure}
    \centering
    \includegraphics[width=1\textwidth]{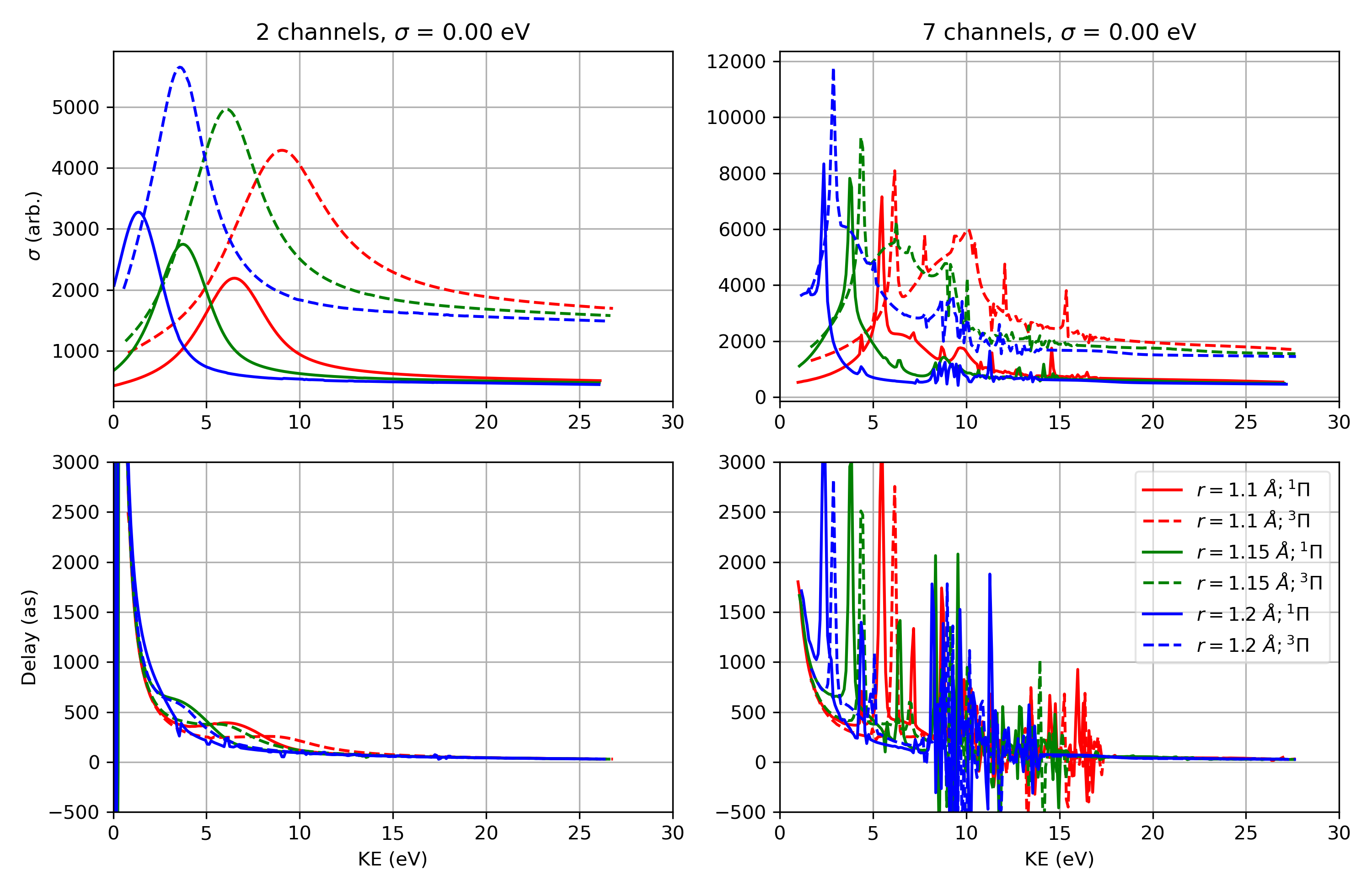}
    \caption{Angle- and orientation-averaged cross-section (top panel) and photoemission delay (bottom panel) for the MCSCI calculations at different internuclear separations.}
    \label{fig:sm:rll-fwhm-0eV}
\end{figure}

\begin{figure}
    \centering
    \includegraphics[width=1\textwidth]{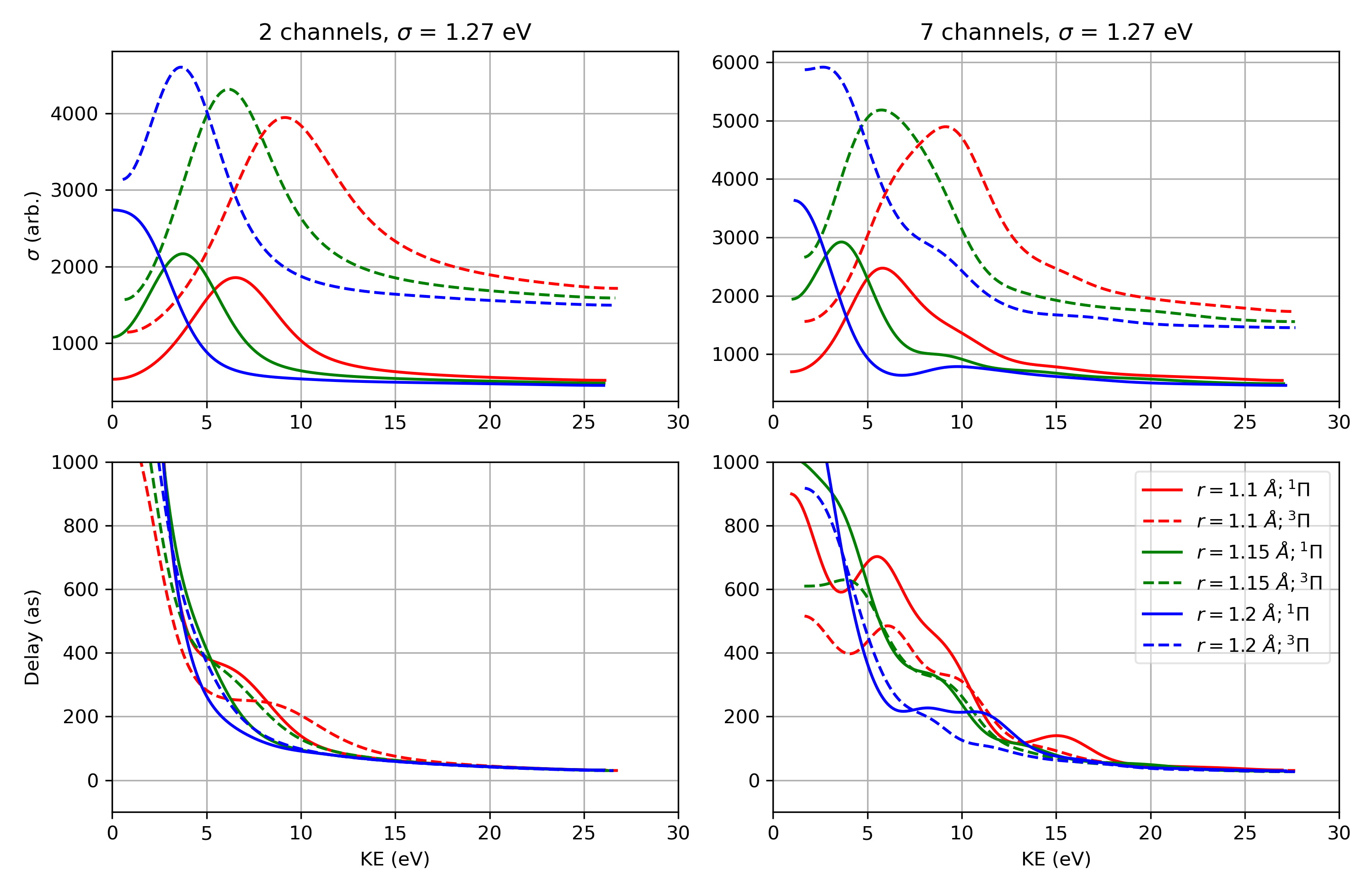}
    \caption{Angle- and orientation-averaged cross-section (top panel) and photoemission delay (bottom panel) for the MCSCI calculations at different internuclear separations. The results have been convolved with a gaussian kernel of $\sigma$=3~eV to account for the photon energy resolution in our measurement}
    \label{fig:sm:rll-fwhm-3eV}
\end{figure}

The MCSCI calculations are performed at three different nuclear geometries: a nuclear separation, $r_0$, of 1.1~\AA,~1.15~\AA~ and 1.2~\AA.
Figure \ref{fig:sm:rll-fwhm-0eV} shows the calculated cross-section (top panels) and angle- and orientation-averaged photoemission delay (bottom panels) for the three different nuclear geometries.
It is notable that a change in bond axis length of only 5 picometers is shown to shift the energetic position of the shape resonance (i.e. the electron kinetic energy at which the shape resonance appears) by several electron volts.
Dashed lines represent the cross-section and delay for ionization to the triplet $\sigma(^3\Pi)$ state of the molecular ion and solid lines to the singlet $\sigma(^1\Pi)$ state.
The cross sections roughly follow the expected statistical behavior of $\frac{\sigma(^3\Pi)}{\sigma(^1\Pi)} \sim3$.
The sharp features visible in the seven-channel calculation correspond to autoionizing resonances accessed through coupling between the different ionization channels.
To account for the finite photon energy resolution in our experiment, we convolve the simulated photoemission delay with a gaussian kernel with a full width at half maximum of $3$~eV).
These convolved results are shown in Fig. \ref{fig:sm:rll-fwhm-3eV}.

\begin{figure}
    \centering
    \includegraphics[width=0.5\textwidth]{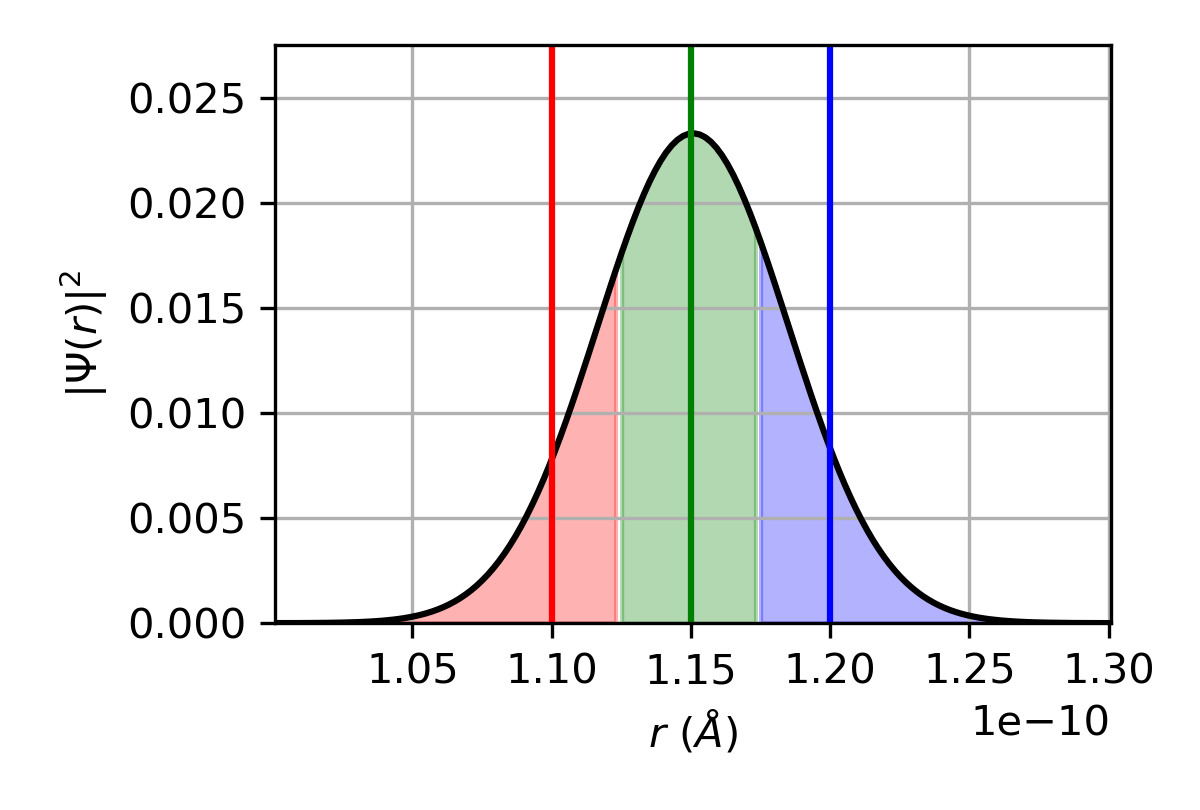}
    \caption{Calculated population of internuclear geometries for ground state harmonic wavefunctino with fundamenta vibrational constant $\omega_e$ of 1904.2~cm$^-1$.}
    \label{fig:sm:nuc_wfn}
\end{figure}

We simulate the overall cross-section and photoemission delay from our measurement by incoherently averaging the calculations for the three different internuclear separations and the singlet and triplet states of the molecular ion.
The contribution from each geometry is weighted by the estimated population of that geometry in the ground vibrational state of the molecule.
The weights are calculated by integrating the probability distribution shown in Fig. \ref{fig:sm:nuc_wfn} across the three independent regions which are closest to each of the three geometries.
This probability distribution was calculated according to an internuclear separation $r_0$ of 1.15077 \AA~ and a fundamental vibrational constant $\omega_e$ of 1904.2~cm$^-1$ \cite{brown_magnetic_1972}.

\section{Photoemission Delay Extraction}
The delay between two photoemission events is directly proportional to the difference in the angular deviation $\Delta\theta$ the corresponding electrons are given by the streaking laser.
In the case of the measurement described here, the oxygen and nitrogen photoelectrons experience a change in final momentum due to the streaking field of $\theta_O,i$ and $\theta_N,i$ at each XFEL shot $i$.
The difference between these two angles, $\theta_O,i - \theta_N,i = \delta\theta$, maps linearly to the photoemission delay:
\begin{equation}
\Delta\tau=\frac{\Delta\theta}{2\pi}\times T_L.
\label{eqn:sm:ang_to_time}
\end{equation}

The measurement of $\Delta\theta$ is complicated by the inherent temporal jitter between the x-ray pulse and the streaking laser, which has an error distribution of width $\sim$500 fs \cite{glownia_time-resolved_2010}.
As a result of this temporal jitter, $\theta_O,i$ and $\theta_N,i$ change across every shot.
This precludes a standard analysis scheme where the reference is held fixed.
However, we can exploit this temporal jitter to extract the difference in streaking angle between the two photolines through correlation.

Our analysis framework is formally analyzed in an upcoming publication and was employed in Ref.~\cite{guo_experimental_2023}.
We summarize the salient details in section~\ref{sec:sm:corr-theory} below.
We define two radial regions on the detector, corresponding to the high momentum portion of each of the two photolines resulting from ionization of the N and O $1s$ orbitals.
The lower limit of each region is defined by the momentum at which the the radial gradient of the 2D projection of the unstreaked electron distribution is maximized.
The upper limit is chosen to ensure all high energy electrons are captured.
The radial position of the maximal gradient depends on the kinetic energy of the electrons, which in turn depends on the incident photon energy.
In this measurement, we do not have access to an independent measurement of the incoming x-ray spectrum.
However, we do perform a single-shot measurement of the energy of the electron beam which produces the x-ray pulse.
The energy of the electron beam is related to the x-ray photon energy through the FEL resonance equation:
formula~\cite{bonifacio_collective_1984}:
\begin{equation}
    \lambda_r = \lambda_u \frac{1+\frac{K_u^2}{2}}{2\gamma^2},
    \label{eqn:sm:feleqn}
\end{equation}
where $\gamma$ is the Lorentz factor of the electron beam, $\lambda_u$ and $K_u$ are the undulator period and strength parameters, and $\lambda_r$ is the wavelength of the XFEL radiation.
The energy of the electrons in the electron beam is not monochromatic and the energy distribution within the electron beam is particularly broad for ESASE operation~\cite{duris_tunable_2020}.
The measurement of the electron beam energy performed at the LCLS is derived from the position of the entire electron bunch in a dispersive section of the electron beamline and represents a measurement of the average energy within the bunch.
This means that the calculation of the x-ray photon energy from this number has an error distribution which we have estimated from previous measurements \cite{li_attosecond_2022,duris_tunable_2020} to be $\sim$3~eV at the photon energies used in this measurement.
We can estimate the average photon energy of a number of different shots with much greater accuracy.

To define the lower limit on a single-shot basis, we first performed a fit of the maximal value of the radial gradient for collections of shots with the streaking laser mistimed, at different central photon energies as determined by the electron beam energy.
On each shot, we estimate the central photon energy using the electron beam energy and Eq. \ref{eqn:sm:feleqn} and use the fit to define the lower limit for both the nitrogen and oxygen photoelectrons.
We radially integrate the electron yield measured in these two regions to produce two angle-resolved vectors $I_N(\theta)$ and $I_O(\theta)$, which we calculate on a shot-to-shot basis.
We calculate the partial correlation maps shown in Fig. 2 \textbf{b} in the main text by calculating the partial correlation coefficient between the radially integrated intensities $I_N$ and $I_O$ at each pair of angles $\theta_i, \theta_j$, according to: 
\begin{align}
\rho(I_N, I_O)&=\frac{\PCov\left(I_N,I_O;P\right)}{ \sqrt{\PCov\left(I_N,I_N;P\right)\PCov\left(I_O,I_O;P\right)}},
\label{eq:sm:pcorr}
\end{align}
where
\begin{align}
\PCov\left(X,Y;I\right) = \Cov\left(X,Y\right) - \frac{\Cov\left(X,I\right)\Cov\left(I,Y\right)}{\Cov\left(I,Y\right)}
\end{align}
and
\begin{align}
\Cov\left(X,Y\right) = \langle XY \rangle - \langle X \rangle \langle Y \rangle.
\end{align}
Here, angular brackets denote the average across multiple XFEL shots.

The partial correlation controls for the single-shot x-ray pulse intensity, since the single-shot intensity is known to flucuate in a SASE FEL.
The pulse intensity is monitored by recording the number of electrons produced by the x-ray pulse as it passes through a dilute gas before the beamline \cite{walter_time-resolved_2022}.
From the partial correlation maps, we identify the offset of the maximum value of the correlation coefficient from the diagonal line described by $\theta_N = \theta_O$, in those regions of the detector where the radial gradient dominates.
As described in section \ref{sec:sm:corr-theory}, this offset corresponds to the angular difference between the momentum shift of the two photolines.
We identify the offset by transforming the axes to ($\theta_O, \pi + \theta_O - \theta_N$) and averaging across $\theta_N$ to produce a one-dimensional trace.
The peak of this trace corresponds to the difference in the angular direction of the momentum shift imparted on the two photolines by the streaking laser.
We identify the peak by finding the root of the derivative of the polynomial which fits this trace $\pm 40~{^\circ}$ about the maximum. 
The quoted error bar is the uncertainty in this root due to the uncertainty in the fit coefficients for the polynomial.

\begin{figure}
    \centering
    \includegraphics[width=0.7\textwidth]{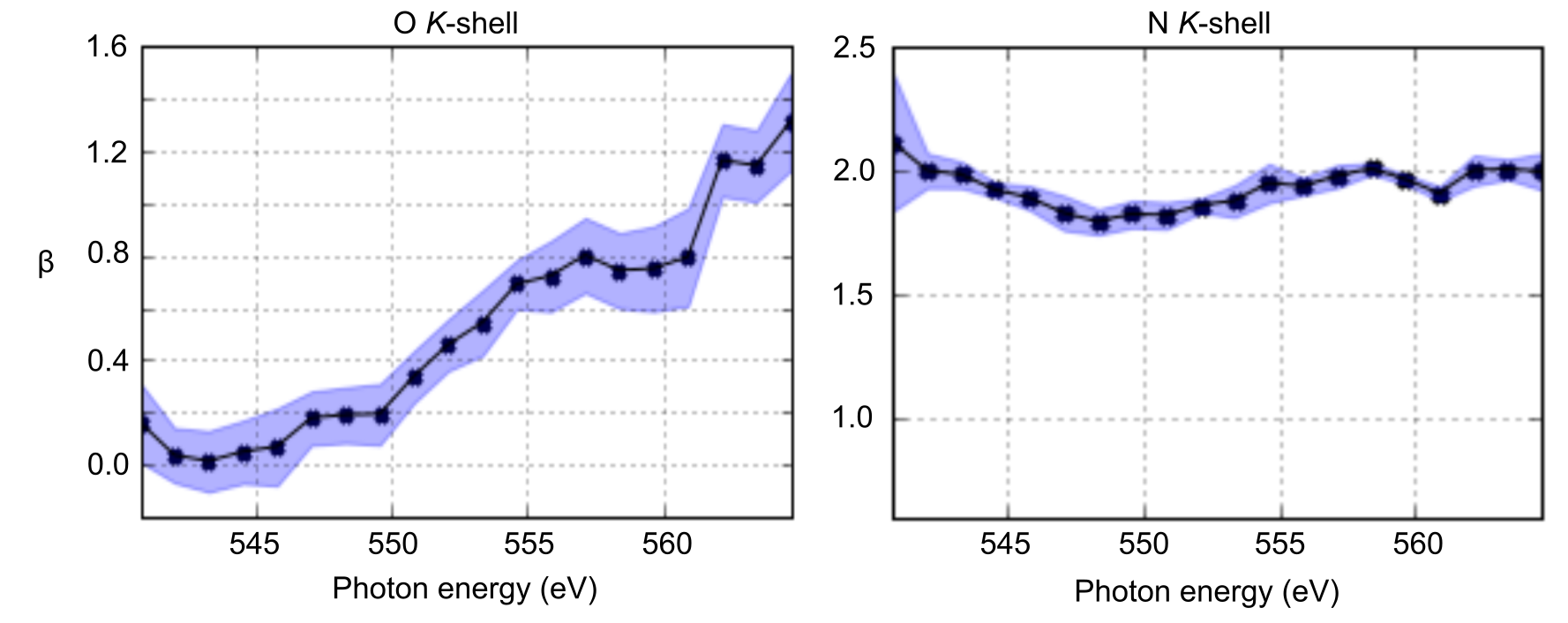}
    \caption{Measured photoelectron asymmetry parameters across the oxygen NEXAFS for nitrogen (left) and oxygen (right) 1$s$ ionization.}
    \label{fig:sm:betas}
\end{figure}

\begin{figure}
    \centering
    \includegraphics[width=1\textwidth]{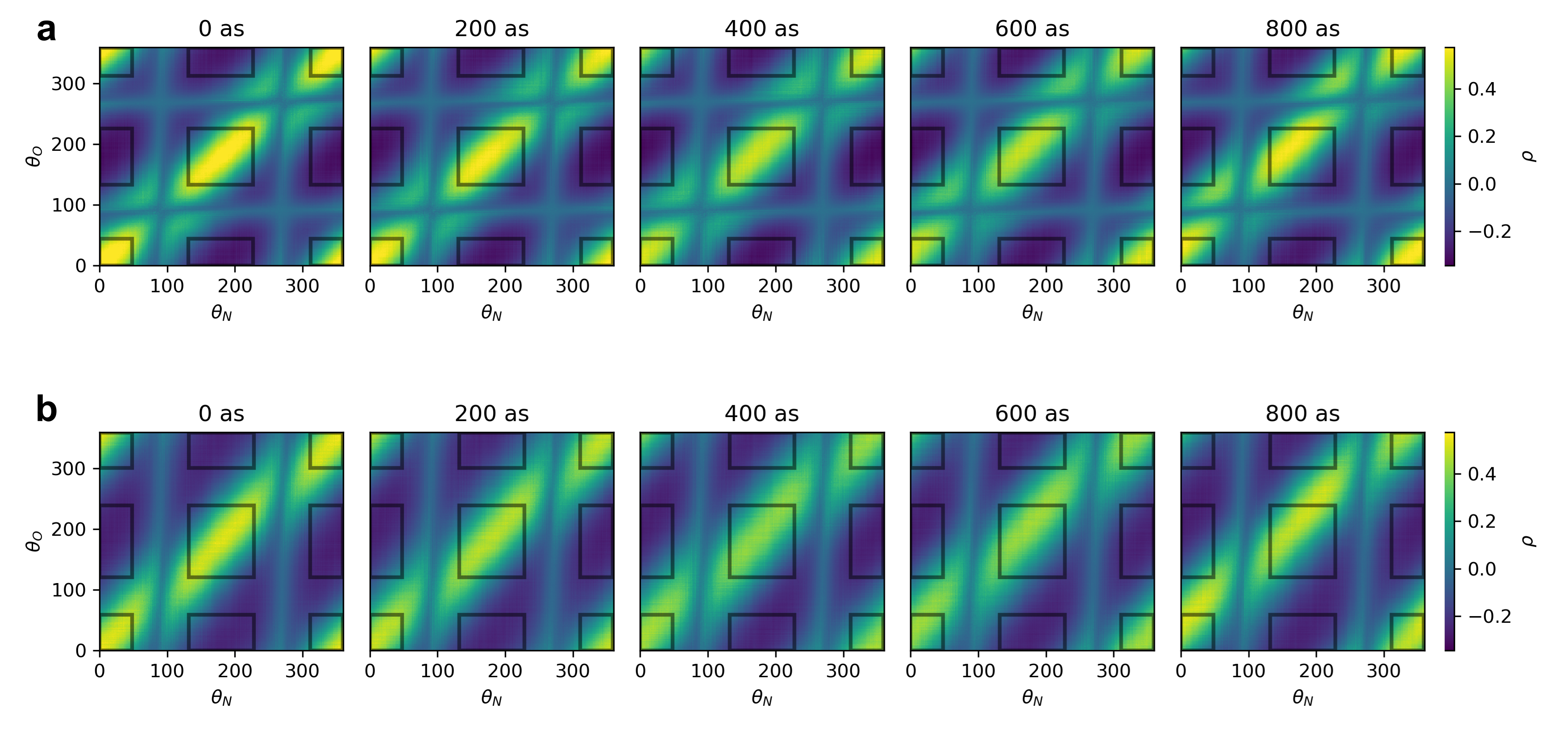}
    \caption{Simulated correlation maps according to Eq. \ref{eq:sm:pcorr} for simulated photoelectron angular distributions of \textbf{a} $\beta(I_N)=\beta(I_O)=2$ and \textbf{b} $\beta(I_N)=2,\beta(I_O)=0$. The offset of the main diagonal feature from the line $I_N=I_O$ increases linearly with the simulated delay $\tau$.}
    \label{fig:sm:sim_maps}
\end{figure}

In addition to the formal theory developed below, we simulate our analysis scheme and verify our method of delay extraction.
Appealing to the equivalence between phase of the photoionization matrix element and a delay measured in angular streaking described in section~\ref{sec:sm:phase-vs-dly}, we simulate the independent photoelectron distributions produced by two 200~as duration x-ray pulses separated by a delay $\Delta\tau$ and the same infrared streaking field.
For five different delay values ($\Delta\tau$ = 0 as, 200 as, 400 as, 600 as, 800 as), we simulate the photoelectron distribution for 10,000 different shots with random phase between the streaking field and the attosecond x-ray pulses.
We identify two regions $I_N(\theta)$ and $I_O(\theta)$ with lower bound at the maximum of the radial gradient of the averaged streaked photoelectron distribution and calculate the correlation maps as defined in Eq. {eq:sm:pcorr}.
The calculated partial correlation maps are shown in Fig. \ref{fig:sm:sim_maps}.
We simulate for two different sets of asymmetry parameters $\beta$: $\beta(I_N)=\beta(I_O)=2$ and $\beta(I_N)=2,\beta(I_O)=0$.
These represent the limiting cases for the photoelectron distributions measured in our experiment (see Fig. \ref{fig:sm:betas}: the N 1$s$ photoline has an asymmetry parameter $\beta=2$ across the spectral region we scan, while asymmetry parameter for the O 1$s$ photoline is measured at $\beta=0$ just above threshold and increases with photon energy).
These values are extracted from XFEL shots where the streaking laser was intentionally mistimed from the x-ray pulse, using the CPBASEX algorithm \cite{cpbasex}.

\begin{figure}
    \centering
    \includegraphics[width=0.7\textwidth]{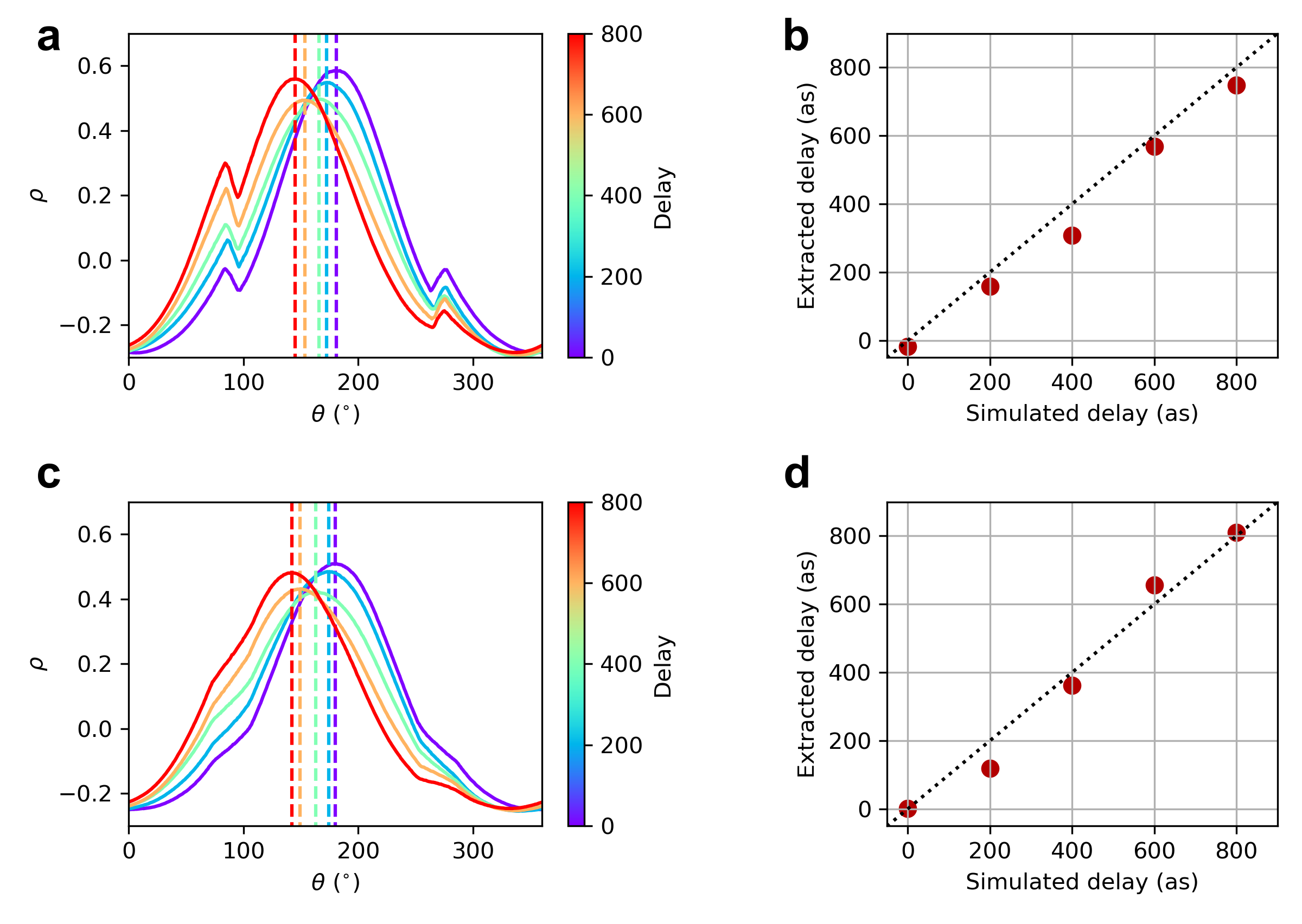}
    \caption{Simulation of extraction of photoemission delay from correlation maps. Panels \textbf{a} and \textbf{c} show the average over $\theta_O$ of the partial correlation maps shown in panels \textbf{a} and \textbf{b} of Fig. respectively. The average has been calculated over the boxes indicated in black in Fig. \ref{fig:sm:sim_maps} and following the coordinate transformation $\theta_N \rightarrow \pi + \theta_O - \theta_N$. Panels \textbf{b} and \textbf{d} show the delay extracted from identifying the peaks of the traces in panels \textbf{a} and \textbf{c}, respectively.}
    \label{fig:sm:delay_extraction}
\end{figure}

As for the experimental data, to extract the delay from these maps as defined in Eq. \ref{eq:sm:pcorr_dly}, we transform the x-axis from $\theta_N$ to $\pi + \theta_O - \theta_N$ and average across $\theta_O$, using regions where the radial gradient dominates.
The resulting traces for the different delays are shown in panels \textbf{a} and \textbf{b} of Fig. \ref{fig:sm:delay_extraction}.
The position of the peak of these traces corresponds to the photoemission delay, as described by Eq. \ref{eq:sm:pcorr_dly}.
As explained above, these simulations indicate that for the inclusion of angles $\pm$60~$^{\circ}$ from the antinodes of the distribution, the additional error accumulated due to neglecting the angular gradient is below $\sim$~10~as at the $2.3$~\textmu m streaking wavelength used in our experiment.
This error grows significantly with the inclusion of angles $\sim70{^\circ}$ from the antinodes.
Therefore, for both simulation and experiment, we use angles which are within 60${^\circ}$ of the x-ray polarization on both photolines.
The simulated delay is compared with the extracted delay in panels \textbf{b} and \textbf{d} of Fig. \ref{fig:sm:delay_extraction}.

\begin{figure}
    \centering
    \includegraphics[width=\textwidth]{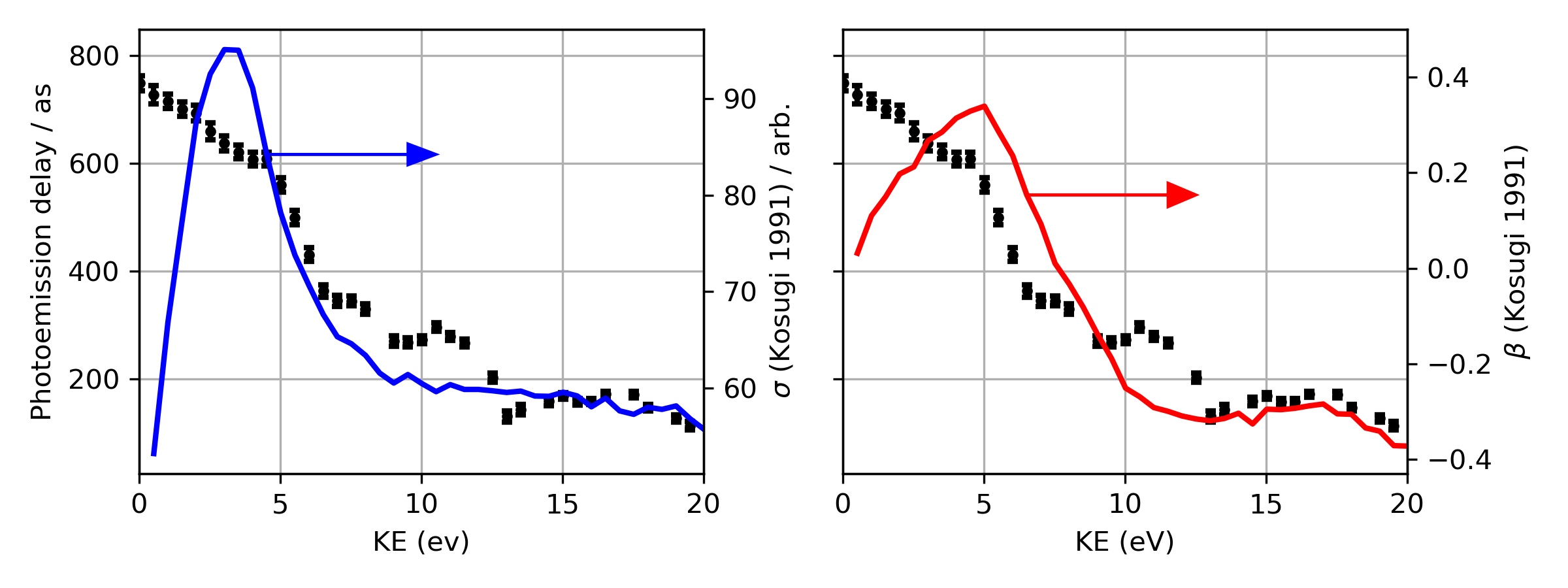}
    \caption{Comparison of the extracted photoemission delay with O $K$-shell ionization $\sigma$ and ionization asymmetry parameter $\beta$ measured by Kosugi \textit{et al.}~\cite{kosugi_highresolution_1992}.}
    \label{fig:sm:kosugi-comp}
\end{figure}

Previous measurements have been reported for the cross-section and angular anisotropy of x-ray ionization of nitric oxide in the oxygen NEXAFS region \cite{kosugi_highresolution_1992}.
We compare these measurements with our extracted delay in Fig. \ref{fig:sm:kosugi-comp}.

\section{Formal Theory of Correlation Analysis}
\label{sec:sm:corr-theory}
The two sets of angle-resolved electron yields, $I_N(\theta)$ and $I_O(\theta)$, correspond to the regional integral of the distribution density $X_N(\vec{p}), X_O(\vec{p})$ where $\vec{p}$ is the two-dimensional VMI projection of the three-dimension photoelectron momentum distribution. 
These densities depend on properties of the photoemission process as well as the vector potential $\vec{A}(t)$.
When the electron emission process is much shorter than the IR cycle, we approximate the distribution of streaked electrons with the displacement of the unstreaked distribution by an instantaneous IR vector potential
\begin{equation}
X_\alpha(\vec{p}; \vec{A}) \simeq X_\alpha(\vec{p} - e\vec{A}(t_X+\tau_\alpha); 0)~,\quad \alpha=N,O~,~\label{eq-model:displacement}
\end{equation}
where $t_X$ is the arrival time of the x-ray pulse, and $\tau_\alpha$ is the additional photoemission delay of each feature.
For simplicity, we denote $\vec{k}_\alpha\equiv e\vec{A}(t_X+\tau_\alpha)$ and $X_{0,\alpha} \equiv X_\alpha(\cdot; 0)$.

Expanding Eq.~(\ref{eq-model:displacement}) w.r.t. the displacement and ignoring the angular gradient, the streaked distribution density can be further approximated as
\begin{equation}
X_\alpha(\vec{p}; \vec{A}) \approx X_{0,\alpha}(\vec{p}) - \frac{\partial X_{0,\alpha}(\vec{p})}{\partial r} \vec{e}_r \cdot \vec{k}_\alpha = X_{0,\alpha}(\vec{p}) - \frac{\partial X_{0,\alpha}(\vec{p})}{\partial r} k \cos(\Theta_\alpha-\theta)~,\label{eq-model:expand}
\end{equation}
where $\vec{e}_r$ is the radial unit vector, $\theta$ the direction of $\vec{p}$, $k$ the magnitude of $\vec{k}_N$ and $\vec{k}_O$, and $\Theta_\alpha=\frac{2\pi}{T_L}(t_X+\tau_\alpha)$ is the direction of $\vec{k}_\alpha$ that is uniformly random across $[0,2\pi)$ due to the arrival time jitter. 
Ignoring the angular gradient requires the dominance of the radial gradient, which for a dipole electron distribution 
(assymetry parameter $\beta$=2) is the case in the vicinity of the antinodes of the distribution, and for an isotropic distribution ($\beta$=0) is the case at every angle.
Following Eq.~(\ref{eq-model:expand}), the regional yields $I_\alpha$ are approximated as $I_\alpha(\theta) \approx I_{0,\alpha}(\theta) - G_{r,\alpha}(\theta) k \cos(\Theta_\alpha-\theta)$, with $G_{r,\alpha}$ being the regional integral of $\partial_r X_{0,\alpha}$. 
For unstreaked distribution that is stable from shot to shot, i.e. $I_{0,\alpha}, G_{r,\alpha}$ are stable, the covariance between electron yields should be
\begin{equation}
\Cov\left(I_N(\theta_N), I_O(\theta_O)\right) = G_{r,N}(\theta_N)G_{r,O}(\theta_O) \frac{k^2}{2} \cos(\theta_N-\theta_O+\Delta\theta)~,
\end{equation}
with $\Delta\theta = \frac{2\pi}{T_L}(\tau_O-\tau_N)$ as defined in the main text Eq.~(2), and the Pearson correlation should be simply $\cos(\theta_N-\theta_O+\Delta\theta)$ because the $G_{r,\alpha}$ cancel out.

In the experiment the unstreaked distributions also fluctuate with the single-shot x-ray pulse properties, so the model of total covariance decomposes~\cite{ross_first_1976} as 
\begin{equation}
\Cov\left(I_N, I_O\right) = \Cov\left(\E{I_N|\vec{k}}, \E{I_O|\vec{k}}\right) + \E{\Cov(I_N,I_O|\vec{k})}~,\label{eq-model:decomp}
\end{equation}
with $I_\alpha$ evaluated at the respective angular bins $\theta_\alpha$. 
The second term in Eq.~(\ref{eq-model:decomp}) is approximately the covariance of unstreaked distributions, which can be most removed by taking partial covariance~\cite{Johnson_Wichern_2007} with respect to the pulse energy, leaving the first term, up to a global scaling factor $s_P$:
\begin{align}
\PCov\left(I_N(\theta_N),I_O(\theta_N);P\right) &\approx \Cov\left(\E{I_N|\vec{k}}, \E{I_O|\vec{k}}\right)s_P\\
&\approx \bar{G}_{r,N}\bar{G}_{r,O}\frac{k^2}{2}\cos(\theta_N-\theta_O+\Delta\theta)s_P~.
\end{align}
Similar to the radial gradients, this global scaling factor is cancelled in the partial correlation calculation:
\begin{subequations}
\begin{align}
\rho(\theta_N,\theta_O;P) &= \frac{\zeta(I_N(\theta_N),I_O(\theta_O)) - \zeta(I_N(\theta_N),P)\zeta(I_O(\theta_O),P)}{\sqrt{(1-\zeta^2(I_N(\theta_N),P))(1-\zeta^2(I_O(\theta_O),P))}}\\
&=\frac{\PCov\left(I_N,I_O;P\right)}{ \sqrt{\PCov\left(I_N,I_N;P\right)\PCov\left(I_O,I_O;P\right)}} \approx \cos(\theta_N-\theta_O+\Delta\theta)~.
\label{eq:sm:pcorr_dly}
\end{align}
\end{subequations}
The result is that, for regions of the detector where the radial gradient dominates the 2D projection of the photoelectron distribution, the numerical modeling indicates that the additional error accumulated due to neglecting the angular gradient is below $\sim$~10~as for the $2.3$~\textmu m streaking wavelength used in this work, for the inclusion of angles $\pm$60~$^{\circ}$ from the antinodes of the distribution.

\begin{figure}
    \centering
    \includegraphics[width=0.7\textwidth]{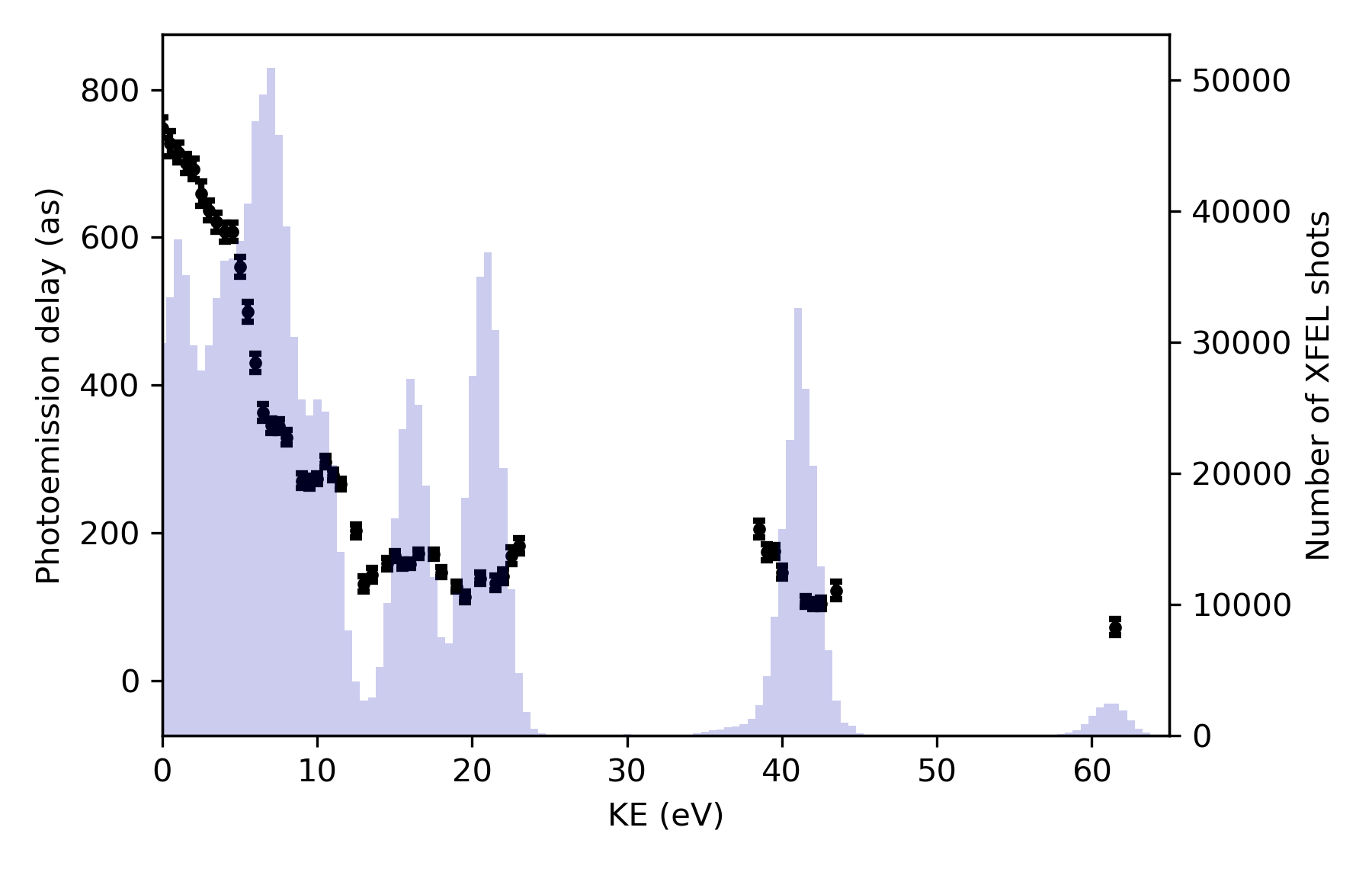}
    \caption{Number of XFEL shots recorded at each central photon energy. In addition to the NEXAFS region of interest examined in this work, it is also possible to record the photoemission delay at $\sim$40~eV and $\sim$60~eV above the oxygen $K$-edge.}
    \label{fig:sm:numshots}
\end{figure}

\section{Experiment}
The geometry of the c-VMI spectrometer differs from a standard VMI spectrometer because the electric field produced by the electrostatic lens plates projects the electrons onto the position-sensitive detector along the same direction as the propagation of the incident laser pulses, as shown in Fig. 1 of the main paper.
A 6.4~mm hole at the center of the MCP allows both the x-ray and infrared light to pass through the detector after the interaction point.
This geometry optimizes the measurement resolution of the photoelectron momentum distribution in the polarization plane of the circularly polarized laser 
pulse~\cite{li_co-axial_2018}.
The thick-lens design of the c-VMI supports detection of electrons of kinetic energies up to a few hundred electron volts, with energy resolution of a few percent.
The c-VMI is installed in the experimental end station inside a 10-inch vacuum chamber. 
The sample was delivered through a pulse supersonic Even-Lavie gas jet, consisting of a 2 mm diameter skimmer (Beam Dynamics, Model 2) $\sim$120 mm from the interaction region. 
The skimmed molecular beam intersects with the x-ray pulse and the streaking laser pulse in the interaction region between the repeller and extractor electrodes in the c-VMI.

\subsection{Streaking Laser}

\begin{figure}
    \centering
    \includegraphics[width=0.7\textwidth]{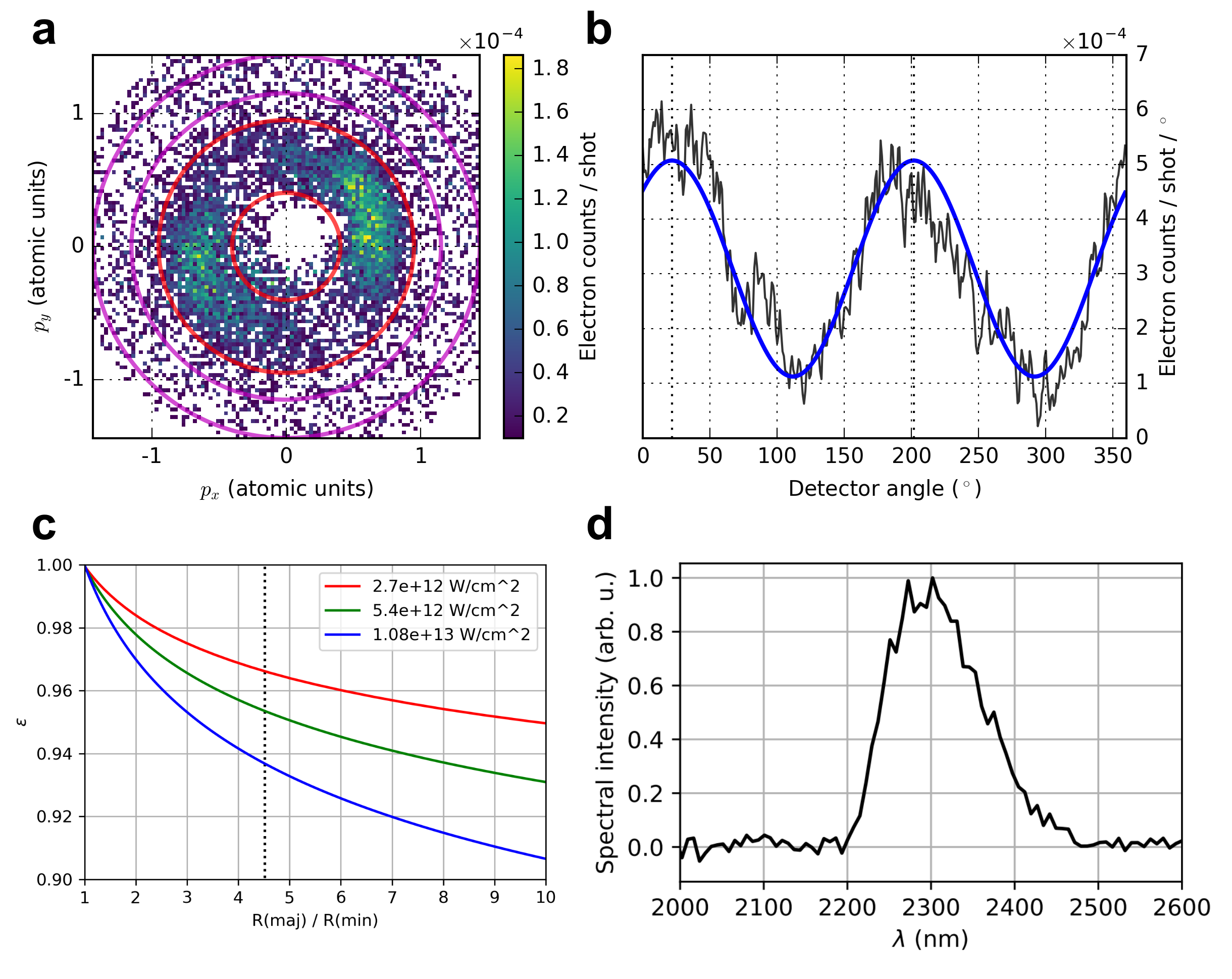}
    \caption{\textbf{a} Above-threshold ionization of nitric oxide by the 2.3~$\mu$m laser field used for the streaking measurement. The region marked by the concentric red circles is radially integrated and this angle-resolved yield is plotted in panel \textbf{b}. The background (marked by the magenta circles) has been subtracted. From this angle-resolved ATI yield, we estimate the contrast ratio between the major and minor axes of the laser polarization in our measurement and determine the ellipticity by simulating this ratio using ADK theory for different laser intensities (panel \textbf{c}). The dashed vertical line shows the ratio extracted from~\textbf{b}. \textbf{d} Streaking laser intensity measured immediately before coupling into the vacuum chamber.}
    \label{fig:sm:laser}
\end{figure}

The IR streaking laser pulse was generated using a commercial optical parametric amplifier~(OPA, TOPAS-HE), pumped with $10$~mJ, $\sim50$~fs, $800$~nm pulses from a Ti:sapphire laser system. 
The resultant $\sim100~\mu$J idler pulses were tuned to be centered at 2.3~$\mu$m, and separated from the signal pulse with a dichroic beamsplitter.
The beam polarization is set by a broad bandwidth quarter-waveplate~(Thorlabs), after which it is focused with an $f=750$~mm lens.
The focused beam is reflected from a dichroic mirror~(HR: 2400 nm / T: visible) before being coupled into the vacuum chamber.
The IR laser is coupled into the interaction point by a holey mirror which the x-rays pass through, allowing the IR beam and the x-ray pulse to co-propagate to the interaction point.
The spectrum of the streaking laser pulse, recorded immediately before it is coupled into the vacuum chamber, is shown in Fig. \ref{fig:sm:laser} \textbf{d}.

We adjusted the intensity of the streaking laser with an iris just before the focusing lens.
We estimate the streaking laser intensity at the interaction point to be $\sim5.4\times10^{12}$W/cm$^{2}$, which was set so that the streaking laser produced $\leq1$ electrons/shot due to above-threshold ionization of the sample.

We characterized the polarization of the streaking laser by increasing the power at the interaction region and characterizing the resultant electron distribution produced by above-threshold ionization.
Figure \ref{fig:sm:laser} \textbf{a} shows the photoelectron spectrum generated by above-threshold ionization~(ATI) of NO molecules using the increased IR intensity and the same quarter-wave plate rotation as in the streaking measurements.

Thanks to the high nonlinearity of the ATI interaction, the angular anisotropy of the resultant photoelectron distribution enables accurate determination of the streaking field ellipticity.
To characterize the polarization, we radially integrate the photoelectron spectrum in the region defined by the red circles in Fig. \ref{fig:sm:laser} \textbf{a} and plot this yield as a function of angle in panel \textbf{d}.
We fit this curve to a sinusoidal curve with varying amplitude, phase offset and DC offset.
This allows us to extract the ratio of the ionization rate along the major and minor axes of the polarization. 
We compare the angular maxima and minima in photoelectron yield with ADK (Ammosov-Delone-Krainov) predictions of tunnel ionization rates \cite{ammasov_tunnel_1986} and identify an ellipticity of $\geq$0.95.

\section{Coulomb Laser Coupling}
The Coulomb potential is long-range and its presence affects the behavior of the electron driven by the streaking field.
For electrons with a asymptotic kinetic energy of several electron volts, it is well understood that the additional contribution to the measured photoemission delay due to this Coulomb-laser interaction can be independently calculated and added to the Wigner delay calculated in section~\ref{sm:ModelPED}.
This is not the case for very low energy electrons, where the terms are no longer straightforwardly separable.
This is a significant contributing factor to the divergence of our calculations from the measured photoemission delay at electron kinetic energies $\leq$2~eV.

This additional delay accrued through the interaction of the electron with the streaking field in the presence of the Coulomb potential is often referred to as the Coulomb-laser coupling (CLC), or continuum-continuum delay.
This delay can be understood as a classical effect and successfully quantified by a classical Monte-Carlo trajectory (CTMC) simulations \cite{pazourek_time-resolved_2013,hofmann_comparison_2013,zimmermann_attosecond_2018}.
It has been well-studied for the case of linear polarization of the streaking laser, but exhibits a strong dependence on the angle between the electric field of the streaking field at the time of ionization and the final position of the electron on the detector, and thus requires careful consideration for the case of a circularly polarized streaking field.
We perform a CTMC simulation to calculate the CLC delay for the streaking laser parameters used in our experiment.
We incorporate this additional delay by adding it to the calculated photoemission delay described above.

\begin{figure}
    \centering
    \includegraphics[width=0.7\textwidth]{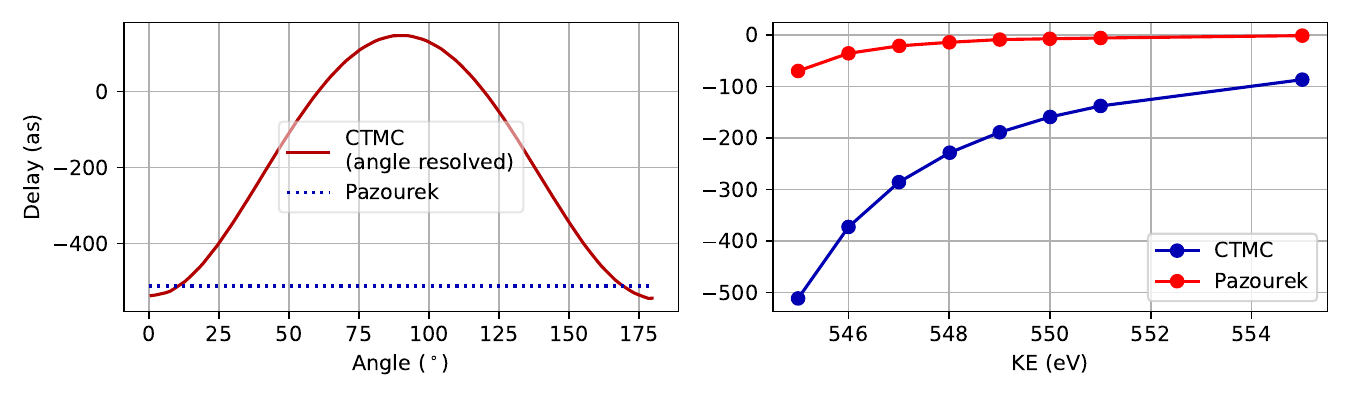}
    \caption{(left) CLC delay as a function of angle between final electron position at detector and electric field at x-ray pulse arrival time. For electrons with trajectory along the polarization axis, the values converge to those of Pazourek \textit{et al.}. (right) angle-integrated delay calculated withing CTMC framework, compared with delay from Pazourek \textit{et al.}}
    \label{fig:sm:clc}
\end{figure}

We can resolve the delay for trajectories with different angles between the laser field at the time of ionization and final position of the electron on the detector.
As shown in Fig. \ref{fig:sm:clc}, for electron trajectories along the polarization axis of the streaking laser, we find the CTMC calculation converges to the analytical result of Pazourek \textit{et al.}~\cite{pazourek_time-resolved_2013}.
However since we average over several emission angles in the measurement, we must calculate a new CLC contribution, which is shown in the right panel of Fig.~\ref{fig:sm:clc}.

\section{Post-Collisional Interaction}

\begin{figure}
    \centering
    \includegraphics[width=0.7\textwidth]{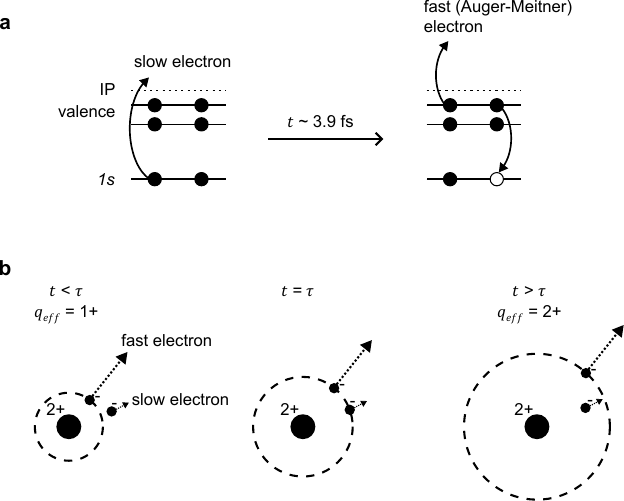}
    \caption{\textbf{a} Schematic of Auger-Meitner decay following oxygen 1$s$ ionization. \textbf{b} For the slow electron, the charge of the doubly charged ion following Auger-Meitner emission is shielded by the Auger-Meitner electron until time $t=\tau$, when the Auger-Meitner electron overtakes the slow electron.}
    \label{fig:smpci-schematic}
\end{figure}

The oxygen $K$-shell photoelectron acquires an additional delay due to its post-collisional interaction with a fast photoelectron produced through Auger-Meitner decay of the core-ionized NO.
This process is illustrated in panel \textbf{a} of Fig. \ref{fig:smpci-schematic}.
X-ray ionization removes an electron from the core 1$s$ orbital of the oxygen atom in the molecule.
In our experiment, this photoionizied electron has energies between $\sim$~1 eV and $\sim$60~eV.
We label this as the `slow' electron.
The core-level vacancy in the molecular cation is highly unstable and it subsequently decays on a short (few femtosecond) timescale.
For light elements, the dominant decay process for a core-ionized system is non-radiative Auger-Meitner decay, where a higher-lying electron in the valence manifold fills the core vacancy and energy is conserved by the emission of a second valence electron.
This emitted electron is the Auger-Meitner electron, and it has kinetic energy equal to the difference between the binding energies of the initial $1s$ orbital and the valence electron which fills the vacancy, less the double ionization potential of the final doubly charged final state.
For oxygen $1s$ ionization, this kinetic energy is around $450$~eV.
We label this Auger-Meitner electron as the `fast' electron.

Panel \textbf{b} depicts the post-collisional interaction between the slow and fast electrons.
The three panels show different times following the emission of the Auger-Meitner electron.
In our experiment, we measure the photoemission delay of the slow electron.
Immediately following the emission of the Auger-Meitner electron, the slow electron is further from the molecular ion because it was emitted before the Auger-Meitner decay took place.
The Auger-Meitner electron provides a charge screening to the slow photoelectron, which is thus emerging from an effective Coulomb potential of charge $q_{eff}$ = 1~+.
This shielding can be approximately described by a Gauss sphere which grows in radius with the fast electron, and is depicted in panel \textbf{b} as a dashed circle.
At time $t=\tau$, the photoelectron and Auger-Meitner electron cross because the Auger electron is moving faster than the photoelectron.
At times $t>\tau$, the Auger-Meitner electron has overtaken the slow photoelectron, which no longer enjoys the screening and thus emerges from an effective Coulomb potential of charge $q_{eff}$ = 2~+.

In some systems, the effect of this change in the potential the electron is emerging from can result in the recapturing of the slow photoelectron by the molecular ion~\cite{eberhardt_photoelectron_1988,berrah_mathitk-shell_2001}.
The interaction also affects the lineshape (i.e. kinetic energy distribution) of Auger-Meitner electrons.
This change in lineshape was successfully calculated for the case of inner shell ionization of argon and xenon by the classical model of Russek and Mehlhorn~\cite{russek_post-collision_1986}.
We use this model to estimate the difference in propagation time between two electrons of the same asymptotic kinetic energy emerging from an ionic potential: both are now traveling in a 2+ Coulomb potential, but one of the electrons began its propagation in a 1+ potential prior to the post-collisional interaction.
To calculate this time difference, we find the time $\tau$ at which the slow electron and the fast Auger-Meitner electron cross, illustrated by the central panel in Fig. \ref{fig:smpci-schematic} \textbf{b}.
This takes place when the two electrons are at the same radius, $\rho$, from the molecular ion.
If the Auger-Meitner electron is emitted at time $t$, the time at which it overtakes the slow electron is
\begin{align}
\tau(t;E_P,E_A) &= \int_{R_i}^{\rho}\frac{dr}{\sqrt{2(E_P+1/r)}} \\
&= t + \int_{R_i}^{\rho}\frac{dr}{\sqrt{2(E_A+2/r)}}
\end{align}
Once the electrons have crossed, the final eKE of the photoelectron drops by $\Delta\simeq 1/\rho$, $E_{Pf}=E_P-\Delta$. 
Therefore, for a given final position $R_f$ and final photoelectron energy $E_{Pf}$, the elapsed time is
\begin{align}
T(E_{Pf},\rho) = \int_{R_i}^{\rho}\frac{dr}{\sqrt{2(E_{Pf}+\Delta+1/r)}} + \int_{\rho}^{R_f}\frac{dr}{\sqrt{2(E_{Pf}+2/r)}}
\end{align}
and for the reference photoelectron that always propagates through a 2+ potential, this time is
$$T(E_{Pf},\rho=0) = \int_{R_i}^{R_f}\frac{dr}{\sqrt{2(E_{Pf}+2/r)}}$$
Therefore, the additional delay accrued by the slow photoelectron due to the post-collisional interaction is
\begin{align}
\Delta T(E_{Pf},\rho) &= \int_{R_i}^{\rho}dr\left(\frac{1}{\sqrt{2(E_{Pf}+1/\rho+1/r)}}-\frac{1}{\sqrt{2(E_{Pf}+2/r)}}\right) \\
&= \left(I(E_{Pf}+\Delta, 1,1/r) - I(E_{Pf}, 2,1/r)\right)\Big|_{r=R_i}^{\rho}
\end{align}

\begin{figure}
    \centering
    \includegraphics[width=0.7\textwidth]{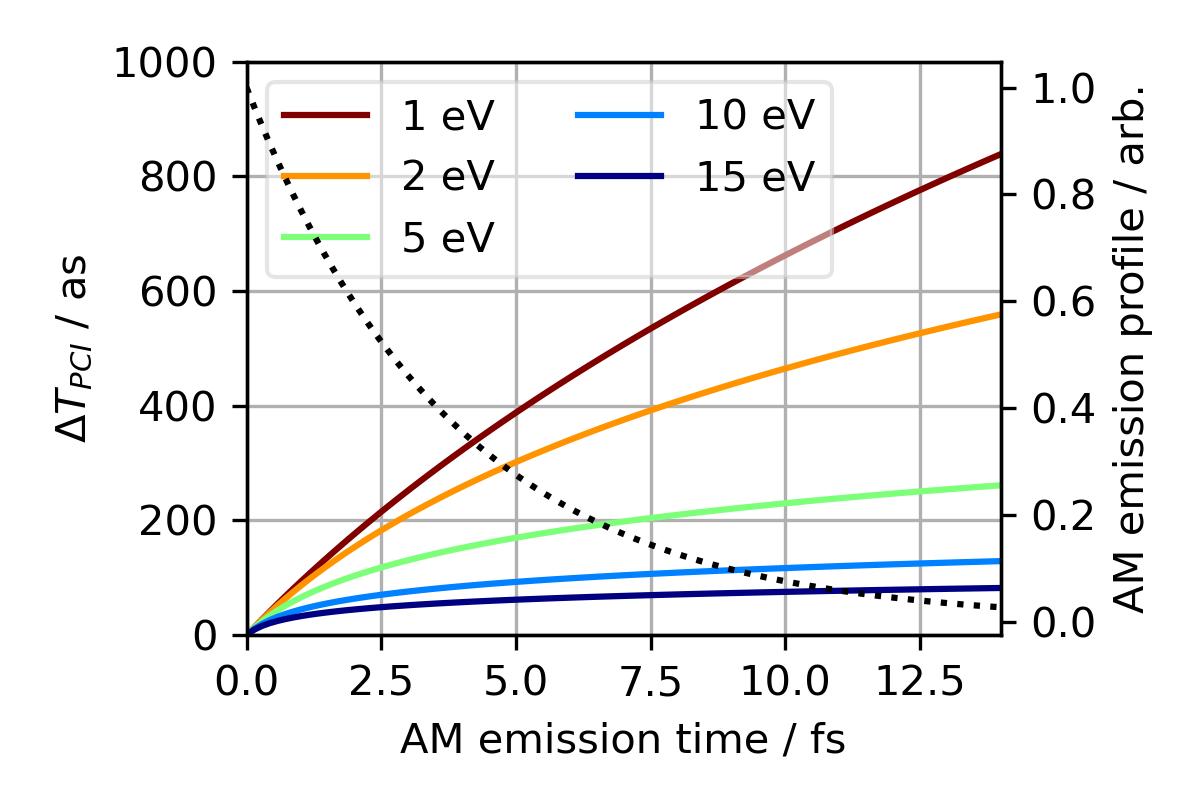}
    \caption{Decay profile for Auger-Meitner decay with decay constant $\Gamma$=3.7~fs (black dotted line). Dependence of additional photoemission delay due to post-collisional interaction on Auger-Meitner emission time, for different asymptotic kinetic energy of the slow photoelectron (solid colored lines).}
    \label{fig:sm:pci-numbers}
\end{figure}

Solving for $\rho(t,E_{Pf},E_A)$ according to
\begin{align}
\int_{R_i}^{\rho}\frac{dr}{\sqrt{2(E_{Pf}+1/\rho+1/r)}} = t + \int_{R_i}^{\rho}\frac{dr}{\sqrt{2(E_A+2/r)}}
\end{align}
and using the following indefinite integral from Eq. (18$b$) in Ref.~\cite{russek_post-collision_1986}:
\begin{align}
I({E}, Z, S)=\frac{({E}+Z S)^{1 / 2}}{2^{1 / 2} {E} S}+\frac{Z}{(2 {E})^{3 / 2}} \ln \left(\frac{(1+Z S / {E})^{1 / 2}-1}{(1+Z S / {E})^{1 / 2}+1}\right)
\end{align}
yields the final additional photoemission delay for a slow photoelectron of asymptotic kinetic energy $E_{Pf}$, a fast Auger-Meitner electron of asymptotic kinetic energy $E_A$ and emission time $t$.
For our measurement, the additional photoemission delay is averaged over different Auger-Meitner emission times corresponding to the decay profile shown in dotted black in Fig. \ref{fig:sm:pci-numbers}, with a decay constant $\Gamma$ of 3.9~fs.
We use an Auger-Meitner electron kinetic energy of 450~eV.
The dependence of the additional photoemission delay on Auger-Meitner emission time for several different slow photoelectron kinetic energies is by the solid colored lines.

\bibliography{referencesJPC.bib, references-additional.bib}

%% file: authors.tex

\author*[1,2]{\fnm{Taran} \sur{Driver}} \email{tdriver@stanford.edu}

\author[3]{\fnm{Miles} \sur{Mountney}}
\author[1,2,4]{\fnm{Jun} \sur{Wang}}
\author[5]{\fnm{Lisa} \sur{Ortmann}}

\author[6]{\fnm{Andre} \sur{ Al-Haddad}}
\author[7]{\fnm{Nora} \sur{Berrah}}
\author[6,8]{\fnm{Christoph} \sur{Bostedt}}
\author[1]{\fnm{Elio~G.} \sur{Champenois}}
\author[5]{\fnm{Louis~F.} \sur{DiMauro}}
\author[9]{\fnm{Joseph} \sur{Duris}}
\author[13]{\fnm{Douglas} \sur{Garratt}}
\author[2]{\fnm{James~M.} \sur{Glownia}}
\author[4,9]{\fnm{Zhaoheng} \sur{Guo}}
\author[10]{\fnm{Daniel} \sur{Haxton}}
\author[1,2,4]{\fnm{Erik} \sur{Isele}}
\author[11]{\fnm{Igor} \sur{Ivanov}}
\author[12]{\fnm{Jiabao} \sur{Ji}}
\author[1]{\fnm{Andrei} \sur{Kamalov}}
\author[9]{\fnm{Siqi} \sur{Li}}
\author[2]{\fnm{Ming-Fu} \sur{Lin}}
\author[13]{\fnm{Jon~P.} \sur{Marangos}}
\author[7]{\fnm{Razib} \sur{Obaid}}
\author[1]{\fnm{Jordan~T.} \sur{O'Neal}}
\author[14,15]{\fnm{Philipp} \sur{Rosenberger}}
\author[2,17]{\fnm{Niranjan~H.} \sur{Shivaram}}
\author[1]{\fnm{Anna~L.} \sur{Wang}}
\author[2]{\fnm{Peter} \sur{Walter}}
\author[1,2]{\fnm{Thomas~J.~A.} \sur{Wolf}}
\author[12]{\fnm{Hans~Jakob} \sur{W\"{o}rner}}
\author[9]{\fnm{Zhen} \sur{Zhang}}
\author[1,4]{\fnm{Philip~H.} \sur{Bucksbaum}}
\author[1,2,14,15]{\fnm{Matthias~F.} \sur{Kling}}
\author[5]{\fnm{Alexandra~S.} \sur{Landsman}} \email{landsman.7@osu.edu}
\author[16]{\fnm{Robert~R.} \sur{Lucchese}} \email{rlucchese@lbl.gov}
\author[3]{\fnm{Agapi} \sur{Emmanouilidou}} \email{a.emmanouilidou@ucl.ac.uk}
\author*[1,9]{\fnm{Agostino} \sur{Marinelli}} \email{marinelli@slac.stanford.edu}
\author*[1,2]{\fnm{James~P.} \sur{Cryan}} \email{jcryan@slac.stanford.ed}


\affil[1]{\orgdiv{Stanford Pulse Institute}, 
    \orgname{SLAC National Accelerator Laboratory}, 
    \orgaddress{\street{2575 Sand Hill Rd.}, \city{Menlo Park} \postcode{94025}, \state{California}, \country{USA}}}
\affil[2]{\orgdiv{Linac Coherent Light Source}, 
    \orgname{SLAC National Accelerator Laboratory}, 
    \orgaddress{\street{2575 Sand Hill Rd.}, \city{Menlo Park} \postcode{94025}, \state{California}, \country{USA}}}
\affil[3]{\orgdiv{Department of Physics and Astronomy}, 
    \orgname{University College London}, \orgaddress{\street{}, \city{London}, \postcode{}, \country{United Kingdom}}}
\affil[4]{\orgdiv{Applied Physics Department}, 
    \orgname{Stanford University}, 
    \orgaddress{\street{}, \city{Stanford}, \postcode{94305}, \state{California}, \country{USA}}}
\affil[5]{\orgdiv{Department of Physics}, 
    \orgname{The Ohio State University}, 
    \orgaddress{\street{}, \city{Columbus}, \postcode{43210} \state{Ohio}, \country{USA}}}
\affil[6]{\orgname{Paul-Scherrer Institute}, 
    \orgaddress{\street{}, \city{Villigen PSI}, \postcode{} \country{Switzerland}}}
\affil[7]{\orgdiv{Department of Physics}, 
    \orgname{University of Connecticut}, 
    \orgaddress{\street{}, \city{Storrs}, \postcode{}, \state{Connecticut}, \country{USA}}}
\affil[8]{\orgdiv{LUXS Laboratory for Ultrafast X-ray Sciences, Institute of Chemical Sciences and Engineering}, 
    \orgname{Ecole Polytechnique Federale de Lausanne (EPFL)},
    \orgaddress{\street{}, \city{Lausanne}, \postcode{}, \country{ Switzerland}}}
\affil[9]{\orgname{SLAC National Accelerator Laboratory}, 
    \orgaddress{\street{2575 Sand Hill Rd.}, \city{Menlo Park} \postcode{94025}, \state{California}, \country{USA}}}
\affil[10]{\orgname{KLA Corporation}, 
    \orgaddress{\city{Milpitas}, \postcode{95035}, \state{California}, \country{USA}}}
\affil[11]{\orgdiv{Center for Relativistic Laser Science}, 
    \orgname{Institute for Basic Science}, 
    \orgaddress{\street{}, \city{Gwangju}, \postcode{61005}, \country{Korea}}}
\affil[12]{\orgdiv{Laboratorium fur Physikalische Chemie}, 
    \orgname{ETH Zurich}, 
    \orgaddress{\street{}, \city{Zurich}, \postcode{}, \country{Switzerland}}}
\affil[13]{\orgdiv{The Blackett Laboratory}, 
    \orgname{Imperial College London}, 
    \orgaddress{\street{}, \city{London}, \postcode{}, \country{United Kingdom}}}
\affil[14]{\orgdiv{Physics Department}, 
    \orgname{Ludwig-Maximilians-Universitat}, 
    \orgaddress{ \street{}, \city{Munich}, \postcode{}, \state{Garching}, \country{Germany}}}
\affil[15]{\orgname{Max Planck Institute of Quantum Optics}, 
    \orgaddress{\street{}, \city{Garching}, \postcode{}, \country{Germany}}}
\affil[16]{\orgdiv{Chemical Sciences Division}, 
    \orgname{Lawerence Berkeley National Laboratory}, 
    \orgaddress{\street{}, \city{Berkeley}, \postcode{}, \state{California}, \country{USA}}}
\affil[17]{\orgdiv{Department of Physics and Astronomy and Purdue Quantum Science and Engineering Institute},
    \orgname{Purdue University}, 
    \orgaddress{\street{}, \city{West Lafayette}, \postcode{}, \state{Indiana}, \country{USA}}}